\documentclass[utf8]{frontiersSCNS} 

\usepackage{url,hyperref,lineno,microtype,subcaption}
\usepackage[onehalfspacing]{setspace}

\def\keyFont{\fontsize{8}{11}\helveticabold }
\def\firstAuthorLast{Segneri {et~al.}} 
\def\Authors{Marco Segneri\,$^{1}$, Hongjie Bi\,$^{1}$, Simona Olmi\,$^{2,3}$, and Alessandro Torcini$^{1,3,*}$}
\def\Address{$^{1}$ Laboratoire de Physique Th\'eorique et Mod\'elisation, Universit\'e de Cergy-Pontoise,CNRS, UMR 8089, 95302 Cergy-Pontoise cedex, France \\
$^{2}$ Inria Sophia Antipolis M\'{e}diterran\'{e}e Research Centre, 2004 Route des Lucioles, 06902 Valbonne, France\\
$^{3}$ CNR - Consiglio Nazionale delle Ricerche - Istituto dei Sistemi Complessi, via Madonna del Piano 10, 50019 Sesto Fiorentino, Italy}
\def\corrAuthor{Alessandro Torcini}
\def\corrEmail{alessandro.torcini@u-cergy.fr}

\begin{document}
\onecolumn
\firstpage{1}

\title[Theta-nested gamma oscillations]{Theta-nested gamma oscillations in next generation neural mass models} 

\author[\firstAuthorLast ]{\Authors} 
\address{} 
\correspondance{} 

\extraAuth{}

\maketitle

\begin{abstract}
Theta-nested gamma oscillations have been reported in many areas of the brain and are believed
to represent a fundamental mechanism to transfer information across spatial and temporal scales.
In a series of recent experiments {\it in vitro} it has been possible to replicate with an optogenetic theta
frequency stimulation several features of cross-frequency coupling (CFC) among theta and gamma rhythms observed 
in behaving animals. In order to reproduce the main findings of these experiments we have considered 
a new class of neural mass models able to reproduce exactly the macroscopic dynamics
of spiking neural networks. In this framework, we have examined two set-ups able
to support collective gamma oscillations: namely, the
pyramidal interneuronal network gamma (PING) and the interneuronal network gamma (ING).
In both set-ups we observe the emergence of theta-nested gamma oscillations by driving
the system with a sinusoidal theta-forcing in proximity of a Hopf bifurcation.
These mixed rhythms display always phase amplitude coupling. However two different types of nested oscillations
can be identified: one characterized by a perfect phase locking between theta and gamma rhythms,
corresponding to an overall periodic behaviour; another one where the locking is 
imperfect and the dynamics is quasi-periodic or even chaotic.
From our analysis it emerges that the locked states are more frequent in the ING set-up.
In agreement with the experiments, we find theta-nested gamma oscillations for forcing frequencies 
in the range [1:10] Hz, whose amplitudes grow proportionally to the forcing one
and which are clearly modulated by the theta phase. Furthermore, analogously to the
experiments, the gamma power and the frequency of the gamma-power peak increase
with the forcing amplitude. At variance with experimental findings, the gamma-power
peak does not shift to higher frequencies by increasing the theta frequency. This effect
can be obtained, in or model, only by incrementing, at the same time, also the noise or the forcing amplitude.
On the basis of our analysis both the PING and ING mechanisms give rise to theta-nested gamma oscillations 
with almost identical features.

\tiny
\keyFont{ \section{Keywords:} Neural oscillations, neural mass models, cross-frequency coupling, hippocampus, quadratic integrate-and-fire neuron, phase-amplitude coupling} 
\end{abstract}

\maketitle

\section{Introduction}

Oscillations in the brain, reflecting the underlying dynamics of neural populations,
have been measured over a broad frequency range \citep{buzsaki2006}. 
Particularly studied are $\gamma$-rhythms (30-120 Hz), due to their ubiquitous presence
in many regions of the brain, irrespectively of the species \citep{buzsaki2012}, and for their 
relevance for cognitive tasks \citep{fries2007} and neuronal diseases \citep{uhlhaas2006,williams2010}.

Inhibitory networks have been shown to represent a fundamental ingredient for the emergence of $\gamma$-oscillations
\citep{bartos2007synaptic,buzsaki2012}. Indeed, inhibition is at the basis of the two mainly
reported mechanisms: pyramidal interneuronal network gamma (PING) and interneuronal network gamma (ING) \citep{tiesinga2009}.
The ING mechanism is observable in purely inhibitory networks in presence of few ingredients:
recurrent connections, a time scale associated to the synaptic $GABA_A$ receptors and an
excitatory drive sufficiently strong to lead the neurons supra-threshold \citep{buzsaki2012}. The collective oscillations (COs)
emerge when a sufficient number of neurons begins to fire within a short time window
and generate almost synchronous inhibitory post-synaptic potentials (IPSPs) in the post-synaptic
interneurons. The inhibited neurons fire again when the IPSPs have sufficiently decayed
and the cycle will repeat. Thus, the main ingredients dictating the frequency of the COs in the
ING set-up are: the kinetics of the IPSPs and the excitatory drive \citep{whittington1995}.
On the othe hand the PING mechanism is related to the presence of an excitatory and an inhibitory population,
then COs emerge whenever the drive on the excitatory neurons is sufficiently strong to induce
an almost synchronous excitatory volley that in turn elicits an inhibitory one. 
The period of the COs is thus determined by the recovery time of the pyramidal neurons 
from the stimulus received from the inhibitory population \citep{wilson1972}. 
A peculiarity of this mechanism, observed both {\it in vivo} and
{\it in vitro} experiments, is that there is a delay between the firing of the pyramidal cells and 
the interneuronal burst \citep{buzsaki2012}.

Gamma oscillations are usually modulated by theta oscillations 
in several part of the brain, with theta frequencies corresponding to 
4-12 Hz in rodents and to 1-4 Hz in humans. Specific examples have been reported 
for the hippocampus of rodents in behaving animals and during rapid eye movement (REM) sleep 
\citep{lisman2005,colgin2009,belluscio2012,pernia2014,colgin2015}, 
for the visual cortex in alert monkeys \citep{whittingstall2009},
for the neocortex in humans \citep{canolty2006} etc.
This is an example of a more general mechanism of cross-frequency coupling (CFC)
between a low and a high frequency rhythm which is believed to have
a functional role in the brain \citep{canolty2010}. In particular,
low frequency rhythms (as the $\theta$ one) are usually involving 
broad brain regions and are entrained to external inputs and/or to cognitive events;
while the high frequency activity (e.g. the $\gamma$-rhythm) reflects local computation activity.
Thus CFC can represent an effective mechanism to transfer information across
spatial and temporal scales \citep{canolty2010, lisman2013}.
Four different types of CFC of interest for electrophysiology
has been listed in \citep{jensen2007}: phase-phase, phase-frequency,
phase-amplitude and amplitude-amplitude couplings (PPC,PFC,PAC and AAC).
Two more types of CFCs have been later added as emerging from the analysis of coupled nonlinear oscillators \citep{witte2008}
and coupled neural mass models \citep{chehelcheraghi2017}:
frequency-frequency and amplitude-frequency coupling (FFC and AFC).
In this paper we will consider $\theta$-nested $\gamma$ oscillations,
and in this context we will analyze PPC, PFC and PAC between $\theta$ and $\gamma$ rhythms.
The most studied CFC mechanism is the PAC, which corresponds to the modification of the amplitude (or power) of $\gamma$-waves
induced by the phase of the $\theta$-oscillations. This mechanism has been reported in the primary visual cortex of anaesthetized macaques
subject to naturalistic visual stimulation \citep{mazzoni2011},
as well as during the formation of new episodic memories in the human hippocampus \citep{lega2016}.
As discussed in \citep{jensen2007}, $\theta$ phase can often modulate both amplitude (PAC)
and frequency (PFC) of the $\gamma$ oscillations, therefore these two mechanisms can occur
at the same time. PPC, which refers to n:m phase locking between $\gamma$ and $\theta$ phase oscillations 
\citep{tass1998}, has been identified in the rodent hippocampus during maze exploration \citep{belluscio2012}.

Out study is mostly motivated by recent optogenetic experiments  
revealing PAC in areas CA1 and CA3 of the hippocampus and in the 
medial enthorinal cortex (MEC) \citep{akam2012, pastoll2013, butler2016,butler2018}.
These experiments have shown that a sinuoidal optogenetic stimulation 
at $\theta$-frequency of the circuits {\it in vitro} is able to reproduce 
several features of $\theta$-nested $\gamma$ oscillations, usually observed in behaving rats \citep{bragin1995}.
All these experiments suggest that inhibition has a key role in generating this cross-frequency rhythms;
however both ING \citep{pastoll2013} and PING \citep{butler2016,butler2018}
mechanism has been invoked to explain locally generated $\gamma$ oscillations.

PING and ING oscillation mechanism have been qualitatively reproduced by employing heuristic
neural mass models  \citep{wilson1972,gerstner}. However these standard firing rate models
do not properly describe the synchronization and desynchronizaton phenomena
occurring in neural populations \citep{devalle2017,coombes2019}. 
Recently a new generation of neural mass models has been designed, which are  
able to exactly reproduce the network dynamics of spiking neurons of class I,
for any degree of synchronization among the neurons \citep{montbrio2015}.
In particular, for purely inhibitory networks, these mean-field models
have been able to reproduce the emergence of COs, observed in the corresponding networks,
without the inclusion of an extra time delay \citep{devalle2017}, as well as the phenomenon of 
event related synchronisation and desynchronisation \citep{coombes2019}.
 
Our main aim is to understand how $\theta$-nested $\gamma$ oscillations
can emerge when a PING or ING mechanism is responsible for the fast oscillations
and which differences can be expected in the population dynamics in the two cases.
Therefore we will consider the new class of neural mass models introduced by Montbri\'{o} et al. \citep{montbrio2015}
in two configurations: namely, a purely inhibitory population (ING set-up) and two coupled excitatory-inhibitory
populations (PING set-up). In both configurations we will examine the
system response to an external sinusoidal $\theta$-drive.

In particular, Section II is devoted to the introduction of different spiking network configurations of
Quadratic Integrate-and-Fire (QIF) neurons able to generate $\gamma$ COs via PING and ING mechanisms and
to the introduction of their corresponding exact neural mass formulations. 
A detailed bifurcation analysis of the neural mass models for the
PING and ING set-ups, in absence of any external forcing, is reported in Section III.
The PAC mechanism is analysed and discussed in Section IV. Firstly, by considering
different types of PAC states : namely, phase locked or unlocked.
Secondly, by comparing our numerical results for PAC dynamics with
experimental findings, reported in \citep{butler2016} and \citep{butler2018},
for the CA1 region of the hippocampus under sinusoidal optogenetic stimulations.
Finally, a discussion of our results and of their implications, as well as of possible
future developments, will be presented in Section V. The results reported
in the paper are mostly devoted to super-critical Hopf bifurcations, however
a specific example of sub-critical Hopf leading to COs is discussed for the 
PING set-up in Appendix A. Further network configurations ensuring
the emergence of COs via PING mechanisms are presented in Appendix B.

\section{Models and Bifurcation analysis}

\begin{figure}[h]
\includegraphics[width=80mm,clip=true]{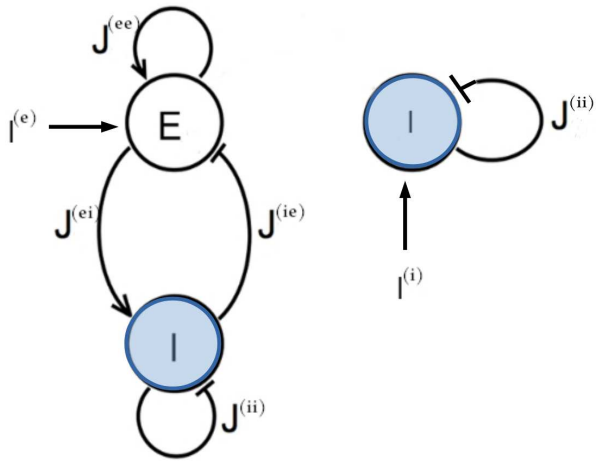}
\caption{\textbf{Network topologies} Two different network configurations have been investigated: on the left side, an excitatory population (E) and
an inhibitory population (I) form a circuit that can generate oscillatory output (PING set-up); on the right side one inhibitory population (I) 
is coupled to itself with an inhibitory coupling (ING set-up).
In both cases an external current $I^{(l)}$ impinging one single population has been considered. 
}
\label{Fig1}
\end{figure}

\subsection{Network Models}

In this paper we want to compare the two principal mechanism at the basis 
of the emergence of collective oscillatory dynamics
in neural networks: namely, the PING and ING mechanisms.
Therefore we will consider QIF neurons in the two following
set-ups: namely, an excitatory and an inhibitory population coupled
via instantaneous synapses (PING configuration) 
or a single inhibitory population interacting via post-synaptic potentials (PSPs) with exponential profile
(ING configuration). The corresponding network configurations are shown in Fig. \ref{Fig1}.
Moreover the neurons are assumed to be fully coupled.  As we will show in the following,
both these two configurations support the emergence of COs.

In particular, the dynamics of the membrane potentials of the QIF neurons in the PING configuration is given by

\begin{eqnarray}
\label{eq:QIF_network}
\nonumber \tau^{(e)}_m \dot{V}^{(e)}_{k} &=& \left(V^{(e)}_{k}\right)^2 + 
\eta^{(e)}_{k} + \tau^{(e)}_m 
\left[J^{(ee)} s^{(e)} - J^{(ie)} s^{(i)}\right] + I^{(e)}(t)
\quad k=1, \dots, N^{(e)}
\\
\nonumber \tau^{(i)}_m \dot{V}^{(i)}_{j} &=& \left(V^{(i)}_{j}\right)^2 + 
\eta^{(i)}_{j} + \tau^{(i)}_m 
\left[J^{(ei)} s^{(e)}-J^{(ii)} s^{(i)} \right] + I^{(i)}(t)
\quad j=1,\dots, N^{(i)}
\\
s^{(l)} &=& \frac{1}{N^{(l)}}\sum_{t^{(l)}_{m}} \delta (t-t^{(l)}_{m}) 
\quad l \in \{e,i\} \qquad ;
\end{eqnarray}
where the super-scripts $e$ ($i$) denotes the excitatory (inhibitory) population, $\tau_m^{(e)}= 20$ ms ($\tau_m^{(i)}= 10$ ms) is
the excitatory (inhibitory) membrane time constant, $\eta^{(l)}_{k}$ is the excitability of the $k$-th neuron 
of population $l$, $J^{(ln)}$ is the strength of the synaptic coupling of population $l$ acting on population $n$. 
The terms $I^{(l)}(t)$ represent a time dependent external current applied to the population $l$;
usually we have considered the external drive to be applied to the excitatory population only, i.e. $I^{(e)}(t) \ne 0$ and $I^{(i)}(t) = 0$.
The synaptic field $s^{(l)}(t)$ is the linear super-position of all the pulses $p(t)$ emitted in the past within the $l$ population, being $p(t)$
a $\delta$-functions in the present case. Furthermore, since the neurons are fully coupled, each neuron will be subject to
the same synaptic field \citep{olmi2010}. The emission of the $m$-th spike in the network occurs at time $t^{(l)}_{m}$ 
whenever the membrane potential of a generic neuron $j$ reaches infinite, i.e. $V^{(l)}_{j}(t^{(l)_-}_{m}) \to +\infty$, 
while the reset mechanism is modeled by setting $V^{(l)}_j(t^{(l)_+}_{j}) \to -\infty$, immediately after the spike emission.\\
The most part of our analysis of the PING set-up will be devoted to networks with self-activation only (i.e. where $J^{(ii)} = 0$), 
a configuration which is known to favour the emergence of collective oscillations \citep{wilson1972, kilpatrick2014, onslow2014}. 
However, as discussed in Appendix B,  we have found that COs can arise in different PING set-ups:
namely, in presence of self-inhibition only (i.e. with $J^{(ii)} \ne 0$ and $J^{(ee)} = 0$)
and in absence of both self-activation and inhibition (i.e. $J^{(ee)} = J^{(ii)} = 0$).

For what concerns the purely inhibitory network, the membrane potential dynamics of the $j$-th neuron is ruled by the
following equations:
\begin{eqnarray}
\label{eq:QIF_network_inh}
\nonumber \tau^{(i)}_m \dot{V}^{(i)}_{j} &=& \left(V^{(i)}_j\right)^2 + 
\eta^{(i)}_{j} - \tau^{(i)}_m J^{(ii)} s^{(i)} + I^{(i)}(t)
\\
\tau_d \dot{s}^{(i)} &=& - {s}^{(i)} + \frac{1}{N^{(i)}}\sum_{t^{(l)}_{m}} \delta (t-t^{(i)}_{m}) 
\quad ;
\end{eqnarray} 
where $\tau_m^{(i)}= 10$ ms. In this case the synaptic field ${s}^{(i)}(t)$
is the super-position of the exponential IPSPs $p(t)= {\rm e}^{-t/\tau_d}/\tau_d$ emitted in the past,
where we set $\tau_d=10$ ms.

For reasons that will be clarified in the next paragraph, we assume that the neuron excitabilities $\eta^{(l)}_{i}$ are randomly distributed 
according to a Lorentzian probability density function  (PDF)
\begin{equation}
\label{eq:lorentzian_eta}
g^{(l)}(\eta) = \frac{1}{\pi}\frac{\Delta^{(l)}}{(\eta-H^{(l)})^2 + (\Delta^{(l)})^2} \;,
\end{equation}
where ${H}^{(l)}$ is the median and $\Delta^{(l)}$ is the half-width half-maximum (HWHM) of the PDF. 
Therefore each population will be composed by neurons supra- (with $\eta^{(l)}_{j} >0$) and sub-threshold (with $\eta^{(l)}_{j} <0$), 
the percentage of one group with respect to the other being determined by the Lorentzian parameters. For the PING set-up we fix $\Delta^{(e)}=\Delta^{(i)} = 1$, 
while varying $H^{(e)}$ and $H^{(i)}$, while for the ING set-up we fix $\Delta^{(i)} = 0.3$
and analyze the dynamics by varying $H^{(i)}$.

The dynamical equations are integrated by employing a 4th order Runge-Kutta
method in absence of noise with a time step $dt = 0.002$ ms ($dt = 0.001$ ms) 
for the PING (ING) set-up. Moreover, we define a threshold $V_p=100$ and a reset value $V_r=-100$.
Whenever the membrane potential $V_j$ of the $j$-th neuron overcomes $V_p$ at a time $t_p$,
it is reset to $V_r$ for a refractory period equal to $2 / V_j$. At the same time the firing time is estimated as $t_p + 1/V_j$;
for more details see \citep{montbrio2015}. The membrane potentials
are initialized from a random flat distribution defined over the range $[-100:100]$,
while the excitabilities are randomly chosen from the Lorentzian distribution \eqref{eq:lorentzian_eta}.
 
In order to characterize the macroscopic dynamics we employ for instantaneous synapses the following indicators:
\begin{equation}
 \label{indicators}
 r^{(l)}(t)=\frac{1}{N^{(l)} \Delta t}\sum_{t^{(l)}_{j}} \delta (t-t^{(l)}_{j}), \quad v^{(l)}(t) = \frac{1}{N^{(l)}}\sum^{N^{(l)}}_{j} V^{(l)}_j(t),
\end{equation}
which represent the average population activity and the average membrane potential
of a population $l$, respectively. In particular the average population activity of the $l-$network $r^{(l)}(t)$ is given 
by the number of spikes emitted in a time window $\Delta t$, divided by the total number of neurons. 
For finite IPSPs we also consider the synaptic field $s^{(l)}(t)$. Furthermore, the emergence of COs in the dynamical evolution, corresponding to 
periodic motions of $r^{(l)}(t)$ and $v^{(l)}(t)$,  are characterized in terms of their frequencies $\nu^{(l)}$.

We assume that the driving current, mimicking the $\theta$-stimulation
in the optogenetic experiments, is a purely sinusoidal excitatory current of the following form \begin{equation}
I_\theta(t) = \frac{I_0}{2} [ 1 - \cos(2\pi \nu_\theta t)]
\label{current}
\end{equation}
where $\nu_\theta$ is the forcing frequency, usually considered within the $\theta$-range, i.e. $\nu_\theta \in [1:10]$ Hz.
In this context a theta phase associated to the forcing field can be defined as $\theta(t) = Mod(2\pi \nu_\theta t, 2 \pi)$. 
For the PING configuration we set $I^{(e)}(t)=I_\theta(t)$ and $I^{(i)}(t) \equiv 0$ and for the ING set-up 
$I^{(i)}(t)=I_\theta(t)$.

\subsection{Neural mass models}

As already mentioned, an exact neural mass model has been derived by Montbri\'{o} et al. \citep{montbrio2015} for a fully coupled network of QIF
neurons with instantaneous synapses and with Lorentzian distributed 
neuronal excitabilities. In this case the macroscopic neural dynamics
of a population $l$ is described by two collective variables: 
the mean field potential $v^{(l)}(t)$ and the instantaneous firing rate $r^{(l)}(t)$.
In this context, the neural mass model for two coupled $E-I$ populations 
with instantaneous synapses, corresponding to the microscopic
model reported in Eq. \eqref{eq:QIF_network}, can be written as 
\begin{eqnarray}
\label{eq:macroscopic_ping}
\dot{r}^{(e)} &=& 
\frac{\Delta^{(e)}}{\left(\tau^{(e)}_m\right)^2 \pi} + 
\frac{2  r^{(e)} v^{(e)}}{\tau^{(e)}_m} \\ 
\nonumber
\dot{v}^{(e)} &=&   
\frac{\left(v^{(e)}\right)^2 +{H}^{(e)} + I^{(e)}(t)}{\tau^{(e)}_m}
- \tau_m^{(e)}\left(\pi r^{(e)}\right)^2 
\nonumber \\
&+& J^{(ee)} r^{(e)} - J^{(ie)} r^{(i)} + A \xi^{(e)}
\nonumber \\
\dot{r}^{(i)} &=& \frac{\Delta^{(i)}}{\left(\tau^{(i)}_m\right)^2 \pi} + 
\frac{2  r^{(i)} v^{(i)}}{\tau^{(i)}_m } 
\nonumber
\\ 
\nonumber
  \dot{v}^{(i)} &=&   
\frac{\left(v^{(i)}\right)^2 + H^{(i)} + I^{(i)}(t)}{\tau^{(i)}_m} 
- \tau^{(i)}_m \left(\pi   r^{(i)}\right)^2 
\nonumber \\
&+& J^{(ei)} r^{(e)} - J^{(ii)} r^{(i)}  + A \xi^{(i)}
\quad .
\nonumber
\end{eqnarray}
In the equations for the evolution of the average membrane potentials we have also inserted an additive noise term of amplitude $A$, 
employed in some of the analysis to mimic the many noise sources present in the brain dynamics.
In particular the noise terms $\xi^{(e)}$ and $\xi^{(i)}$ are
both $\delta$-correlated and uniformly distributed in the interval
$[-1:1]$.

In case of finite synapses, the exact derivation of the corresponding
neural mass model is still feasible for QIF neurons, but
the macroscopic evolution now contains further equations describing
the dynamics of the synaptic field characterizing the considered synapses
\citep{devalle2017,coombes2019}. In particular, for a single inhibitory population 
with exponential synapses, the corresponding neural mass model reads as: 
\begin{eqnarray}
\label{eq:macroscopic_inh}
\dot{r}^{(i)} &=& \frac{\Delta^{(i)}}{(\tau_m^{(i)})^2 \pi} 
+ \frac{2 r^{(i)}v^{(i)}}{\tau_m^{(i)}}\\
\nonumber \dot{v}^{(i)} &=& 
\frac{(v^{(i)})^2 + H^{(i)} + I^{(i)}(t)}{\tau_m^{(i)}} 
-\tau_m^{(i)}(\pi r^{(i)})^2 \\
\nonumber &-& J^{(ii)} s^{(i)} + A \xi^{(i)} \\
\nonumber 	\dot{s}^{(i)} &=& \frac{1}{\tau_{d}}[-s^{(i)}+r^{(i)}].
\end{eqnarray}
In the present case the equation for the average membrane potential
contains, as already shown before in Eqs. (\ref{eq:macroscopic_ping}), an additive noise term of amplitude $A$.

To analyse the stability of the macroscopic solutions of Eqs. \eqref{eq:macroscopic_ping} and
\eqref{eq:macroscopic_inh}, one should estimate the corresponding Lyapunov spectrum
\citep{lyapunov2016}. This can be done by considering the time evolution of the 
tangent vector, which for the PING set-up turns out to be four dimensional, i.e.
${\bf \delta} = \left\{\delta {r}^{(e)}, \delta v^{(e)}, \delta r^{(i)} \delta v^{(i)} \right\}$.
The dynamics of the tangent vector is ruled by the linearization of the Eqs. \eqref{eq:macroscopic_ping}, namely
\begin{eqnarray}
\label{eq:lin_macroscopic_ping}
\delta \dot{r}^{(e)} &=& 
\frac{2  \left(r^{(e)} \delta v^{(e)} + v^{(e)} \delta r^{(e)}\right) }{\tau^{(e)}_m} \\ 
\nonumber
\delta \dot{v}^{(e)} &=&   
\frac{2 v^{(e)} \delta v^{(e)} }{\tau^{(e)}_m}
- 2 \tau_m^{(e)} \pi^2 r^{(e)} \delta r^{(e)}
+ J^{(ee)} \delta r^{(e)} - J^{(ie)} \delta r^{(i)}  
\nonumber \\
\delta \dot{r}^{(i)} &=& \frac{2  \left( r^{(i)} \delta v^{(i)} + v^{(i)} \delta r^{(i)} \right) }{\tau^{(i)}_m } 
\nonumber
\\ 
\nonumber
 \delta \dot{v}^{(i)} &=&   
\frac{2 v^{(i)} \delta v^{(i)}  }{\tau^{(i)}_m} 
- 2 \tau^{(i)}_m \pi^2   r^{(i)} \delta  r^{(i)}
+ J^{(ei)} \delta r^{(e)} - J^{(ii)} \delta r^{(i)} 
\quad .
\nonumber
\end{eqnarray}

For the ING set-up the tangent vector is three dimensional, namely ${\bf \delta} = \left\{\delta r^{(i)}, \delta v^{(i)}, \delta {s}^{(i)} \right\}$, and its time evolution can be obtained
by the linearization of the Eqs. \eqref{eq:macroscopic_inh}, which reads as
\begin{eqnarray}
\label{eq:lin_macroscopic_inh}
\delta \dot{r}^{(i)} &=& \frac{2 \left(r^{(i)} \delta v^{(i)} + v^{(i)} \delta r^{(i)} \right)}{\tau_m^{(i)}}\\
\nonumber \delta \dot{v}^{(i)} &=& 
\frac{2 v^{(i)} \delta v^{(i)} }{\tau_m^{(i)}} 
-2 \tau_m^{(i)} \pi^2 r^{(i)} \delta r^{(i)} 
- J^{(ii)} \delta s^{(i)}   \\
\nonumber \delta	\dot{s}^{(i)} &=& \frac{1}{\tau_{d}}[-\delta s^{(i)}+ \delta r^{(i)}].
\end{eqnarray}

The LS is composed by 4 (3) Lyapunov exponents (LEs) $\left\{\lambda_i  \right\}$
for the PING (ING) set-ups, which quantify the average growth rates of
infinitesimal perturbations along the orthogonal manifolds. In details, LEs are estimated as follows
\begin{equation}
\lambda_i = \lim_{t \to \infty} \frac{1}{t} \log{\frac{|{\bf \delta}(t)|}{|{\bf \delta}(0)|}}
\quad ,
\end{equation}
by employing the technique described in \citep{benettin1980} to maintain the tangent vectors orthonormal
during the evolution. The autonomous system will be chaotic for $\lambda_1 > 0$, while a periodic (quasi-perodic) dynamics will be characterized by $\lambda_1=0$ ($\lambda_1=\lambda_2=0$) and a fixed point by $\lambda_1 <0$. In a non-autonomous system in presence of an external forcing, one Lyapunov exponent will be necessarily zero,
therefore a periodic behaviour corresponds to $\lambda_1 <0$ while a quasi-periodic one to $\lambda_1  =0$ \citep{lyapunov2016}.
 
In absence of noise, neural mass models have been directly integrated by employing a Runge-Kutta 4th order integration scheme, while in presence of
additive noise with a Heun scheme. In both cases the employed time step corresponds to $dt=0.01$ ms.
In order to estimate the Lyapunov spectra we have integrated the direct and tangent
space evolution with a  Runge-Kutta 4th order integration scheme with $dt=0.001$ ms,
for a duration of 200 s, after discarding a transient of 10 s.

Besides LEs, in order to characterize the macroscopic dynamics of the model, we have
estimated the frequency power spectra $P_S^{(e)}(F)$ ($P_S^{(i)}(F)$) of the mean excitatory (inhibitory) membrane
potential $v^{(e)}(t)$ ($v^{(i)}(t)$) for the PING (ING) set-up. 
The power spectra have been obtained calculating the temporal Fourier transform of the mean membrane 
potentials sampled at time intervals of $2$ ms. In the deterministic (noisy) case, time traces
composed of $2048$ ($1024$) consecutive intervals have been considered to estimate the spectra, which are obtained at a
a frequency resolution of $\Delta F=0.244$ Hz ($\Delta F=0.488$ Hz).
Finally, the power spectra have been averaged over 12 (488) independent realizations
for the deterministic (noisy) dynamics. To compare our numerical results with the experimental ones reported in \citep{butler2016}, 
as a measure of the power of the $\gamma$ oscillations, we have estimated the area of the power spectrum $P_\gamma$ 
in an interval $\pm 15$ Hz around the main peak position $F_r$ of the corresponding power spectrum.

\section{Dynamics in absence of forcing}

Due to the low dimensionality of the neural mass models we have been able to obtain the corresponding
bifurcation diagrams by employing the software MATCONT developed for orbit continuation \citep{matcont}.

In particular, we have derived the bifurcation diagrams in absence of forcing ($I^{(e)}=I^{(i)} \equiv 0$) as a function of the median 
${H}^{(e)}$ and ${H}^{(i)}$ (${H}^{(i)}$) of the excitability distributions for the PING (ING) configuration.
In general, we observe either asynchronous dynamics, corresponding to a stable fixed point (a focus) of the neural mass equations, or COs, 
corresponding to stable limit cycles for the same set of equations. 

\subsection{PING set-up}

\begin{figure*}[ht]
\begin{center}
\includegraphics*[width=0.45\textwidth,clip=true]{fig2a.eps}
\includegraphics*[width=0.45\textwidth,clip=true]{fig2b.eps}
\includegraphics*[width=0.45\textwidth,clip=true]{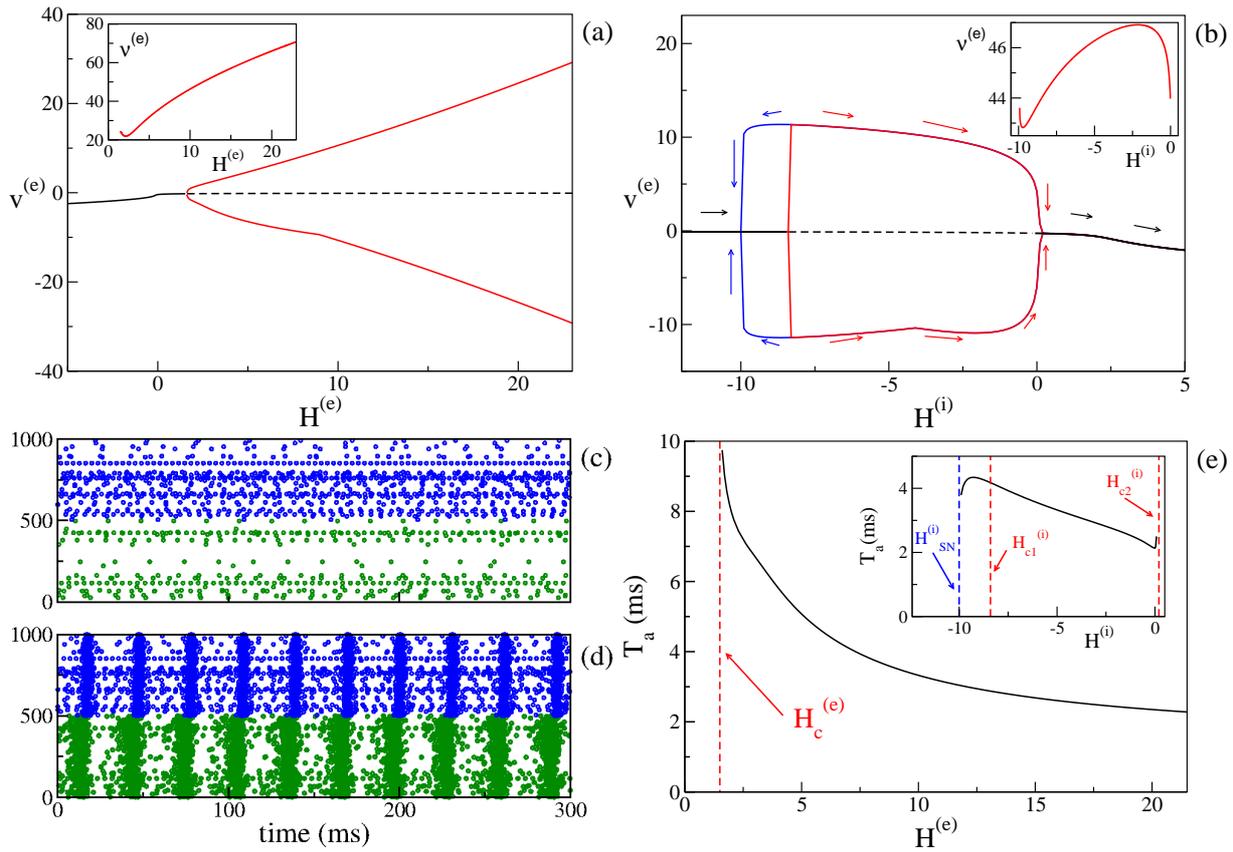}
\includegraphics*[width=0.45\textwidth,clip=true]{fig2d.eps}
\end{center}
\caption {({\bf PING set-up}) (a) Bifurcation diagram of the average membrane potential $v^{(e)}$ as a function of $H^{(e)}$
for $H^{(i)}=-5.0$. The black continuous (dashed) line identifies the stable (unstable) fixed point. The red lines denote the extrema of the limit cycles. The supercritical Hopf bifurcation occurs at $H^{(e)}_c = 1.5$. In the inset is reported the frequency $\nu^{(e)}$ of the COs versus $H^{(e)}$. (b) Bifurcation diagram of the average membrane potential $v^{(e)}$ versus $H^{(i)}$ for $H^{(e)}=10$. The Hopf bifurcations are located at $H_{c1}^{(i)}=-8.4$ and $H_{c2}^{(i)}=0.20$, while the saddle-node of limit cycles
at $H_{SN}^{(i)} = -10.0$. In the inset is reported the frequency $\nu^{(i)} \equiv \nu^{(e)}$ of the COs versus $H^{(i)}$.
(c-d) Raster plots of the excitatory (green dots) and inhibitory (blue dots) networks are calculated in correspondence of the stable fixed point $H^{(e)} = -5.0$ (c) and of the limit cycle $H^{(e)} = +5.0$ (d) for the case analyzed in (a). For a better visualization, the activity of only 500 neurons of each population is shown. (e) Delay $T_a$ as a function of $H^{(e)}$. The red dashed line denotes $H^{(e)}_c$. Here we have used the same parameters as in panel (a). In the inset is reported the dependence of $T_a$ versus $H^{(i)}$ for the parameters
in panel (b). Other parameters of the system: $J^{(ee)} = 8$, $J^{(ie)}=J^{(ei)} = 10$, $J^{(ii)} = 0$ and sizes
of the networks $N^{(e)}$ = 5000, $N^{(i)}$ = 5000.
}
\label{Fig3}
\end{figure*}

For the excitatory-inhibitory set-up, as already mentioned, we usually fix $H^{(i)}=-5$ and we vary $H^{(e)}$. In this case
the inhibitory neurons are mostly below threshold (apart a 6-7 \% of them) and they can be 
driven supra-threshold from the activity of the excitatory population for sufficiently large values of $H^{(e)}$.
COs emerge when a sufficient number of neurons is supra-threshold, i.e. when ${H}^{(e)}$ becomes
sufficiently positive.
Indeed, as shown in Fig. \ref{Fig3} (a), at negative or low values of ${H}^{(e)}$ one has asynchronous dynamics where the neurons
fire indipendently an no collective behaviour is observable (as an example see Fig. \ref{Fig3} (c)).  By increasing  $H^{(e)}$
a supercritical Hopf bifurcation occurs at ${H}^{(e)}_c \simeq 1.5$ leading to the emergence of COs. The COs regime is characterized in the network 
by almost periodic population bursts, where the neurons in one population partially synchronize 
over a short time window of the order of few milliseconds. 
An example for ${H}^{(e)} = 5$ is shown in Fig. \ref{Fig3} (d), where one can observe two salient characteristics of the oscillatory
dynamics. Firstly, the excitatory burst anticipates always the inhibitory one of a certain time interval $T_a$ (in this case $T_a \simeq 5$ ms),
as usually observed for the PING mechanism \citep{tiesinga2009}.
Secondly, the bursts of the excitatory population have a temporal width ($\simeq 8$ ms) which is two or three times larger than those of the 
inhibitory ones ($\simeq 2-3$ ms). This is also due to the fact that a large part of the inhibitory neurons is sub-threshold, therefore
most of them fire within a short time window, irrespectively of their excitabilities, due to the arrival of the synaptic
stimulation from the excitatory population. Instead, the excitatory neurons, which are mostly supra-threshold, recover from silence,
due to the inhibitory stimulation received during the inhibitory burst, over a wider time interval
driven by their own excitabilities. It is evident that the COs frequency of the excitatory and inhibitory population coincide
in this set-up.

Moreover it is important to investigate the bifurcation diagram of the system at fixed median excitatory drive by varying ${H}^{(i)}$.
The corresponding bifurcation diagram is reported in Fig. \ref{Fig3} (b) for ${H}^{(e)} = 10$. By increasing ${H}^{(i)}$
COs emerge from the asynchronous state via a sub-critical Hopf bifurcation at ${H}^{(i)}_{c1} \simeq -8.4$
and they disappear via a super-critical Hopf at ${H}^{(i)}_{c2} \simeq 0.20$.
Since the first transition is hysteretical, COs disappear via a saddle-node of the limit cycles
at a value ${H}^{(i)}_{SN} \simeq -10.00$ lower than ${H}^{(i)}_{c1}$.
Indeed, in the interval $[{H}^{(i)}_{SN}; {H}^{(i)}_{c1}]$ we have the coexistence of a stable focus with a stable limit cycle. In summary, 
COs are clearly observable as long as ${H}^{(i)}$ is negative or sufficiently small. If the inhibitory neurons become mostly supra-threshold, this destroys 
the collective behaviour associated to the PING mechanism.

It is worth noticing that the frequencies of the COs are in the $\gamma$-range, namely $\nu^{(e)} \in [22:71]$ Hz 
(as shown in the inset of Fig. \ref{Fig3} (a)): in this set-up the maximal achievable frequency $\simeq 100$ Hz, since
the decay time of inhibition is dictated by $\tau_m^{(i)} =10$ ms \citep{tiesinga2009}.
On the other hand, the influence of ${H}^{(i)}$ on the frequency of the COs is quite limited. As shown in the inset of Fig. \ref{Fig3} (b) for
a specific case corresponding to ${H}^{(e)} = 10.0$, $\nu^{(i)} \equiv \nu^{(e)}$ varies of few Hz (namely, from $42.8$ to $46.9$ Hz)
by varying ${H}^{(i)}$ of an order of magnitude. 

For what concerns the delay $T_a$ between the excitatory and inhibitory bursts, we observe a decrease of 
$T_a$ with the increase of the excitatory drive ${H}^{(e)}$, from $T_a \simeq 10$ ms
at the Hopf bifurcation, towards 2 ms for large ${H}^{(e)}$ value, see Fig. \ref{Fig3} (e). 
The largest value of $T_a$ is of the order of $\tau_m^{(i)}$. This can be explained
by the fact that the excitatory stimulations should reach the inhibitory population
within a time interval (at most) of $\simeq \tau_m^{(i)}$ to be able to sum up in an effective manner and to ignite the inhibitory burst. 
As shown in the inset of Fig. \ref{Fig3} (e), the increase of ${H}^{(i)}$ has in general the effect to reduce $T_a$; this should be expected since for larger
excitabilities (larger ${H}^{(i)}$), the inhibitory neurons are faster in responding to the excitatory stimulations.
However, this is not the case in proximity of the saddle-node bifurcation at ${H}^{(i)}_{SN}$ and for positive ${H}^{(i)}$,
where the effect is reversed and $T_a$ increases with ${H}^{(i)}$.

For the PING set-up we can observe also sub-crtical Hopf bifurcations: a specific example is discussed in Appendix A in some details.

\subsection{ING set-up}

\begin{figure*}[ht]
\includegraphics*[width=80mm,clip=true]{fig3a.eps}
\includegraphics*[width=80mm,clip=true]{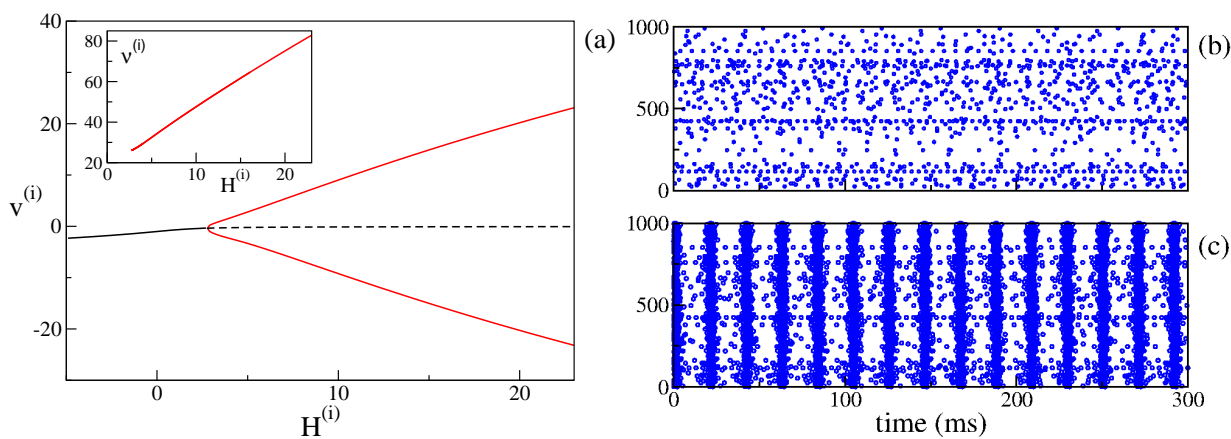}
\caption {({\bf ING set-up}) (a) Bifurcation diagram of the average membrane potential $v^{(i)}$ as a function of $H^{(i)}$. 
The black continuous (dashed) line identifies the stable (unstable) fixed point. The red lines denote the extrema of the limit cycles. 
The supercritical Hopf bifurcation occurs at $H^{(i)}_c \simeq 2.4$. 
In the inset is reported the COs' frequency $\nu^{(i)}$ of the inhibitory population as a function of $H^{(i)}$. 
Right panels: raster plots of the inhibitory network (blue dots) are calculated in correspondence of the stable fixed point $H^{(i)} = 0.0$ (b) and of the limit cycle $H^{(i)} = +10.0$ (c). Only the firing activity of 1000 neurons is displayed.
Parameters of the system: $J^{(ii)} = 21.0$, $\bar{\eta}^{(i)} = 2.0$, $\Delta^{(i)} = 0.3$, $\tau_m^{(i)} = 10.0$ ms, $\tau_d = 10.0$ ms, $A=0$. 
System size for the purely inhibitory network $N^{(i)} = 10000$. 
}
\label{Fig4}
\end{figure*}

As shown in \citep{devalle2017}, in order to observe COs in globally coupled inhibitory QIF networks
and in the corresponding neural mass models, it is sufficient to include a finite synaptic time scale $\tau_d$. 
On the other hand, in sparse balanced QIF networks, COs are observable even for instantaneous synapses \citep{matteo}.
Indeed, for the chosen set of parameters, by varying the median of the inhibitory excitabilities ${H}^{(i)}$, we observe
a super-critical Hopf bifurcation at $H^{(i)}_c \simeq 2.4$, from an asynchronous state to a COs
(see Fig. \ref{Fig4} (a)).
Analogously to the PING set-up, the frequencies of the COs observable in the ING set-up are whithin the $\gamma$-range, namely 
$\nu^{(i)} \in [26:83]$ Hz. In particular, we observe an almost linear increases of $\nu^{(i)}$ with $H^{(i)}$.

Therefore, the PING and ING set-ups here considered are ideal candidates to analyse the
influence of $\theta$-forcing on $\gamma$-oscillatory populations, which represents the main focus of this paper.
In particular, the response of the system to the excitatory $\theta$-forcing current \eqref{current} can be interpreted 
in terms of the bifurcation diagrams for the model in absence of forcing shown, respectively, in Fig. \ref{Fig3} (a)
for the PING set-up  and in Fig. \ref{Fig4} (a) for the ING. The interpretation is possible due to the fact that the response of the system
to the sinusoidal current \eqref{current} can be considered as almost-adiabatic, being the
forcing frequencies $\nu_\theta \in [1:10]$ Hz definitely slower than those of the COs
($\nu^{(e)}$ and $\nu^{(i)}$), which lay in the $\gamma$-range.
 
\section{Dynamics under $\theta$-forcing}

\begin{figure*}[ht]
\begin{center}
\includegraphics*[width=0.45\textwidth,clip=true]{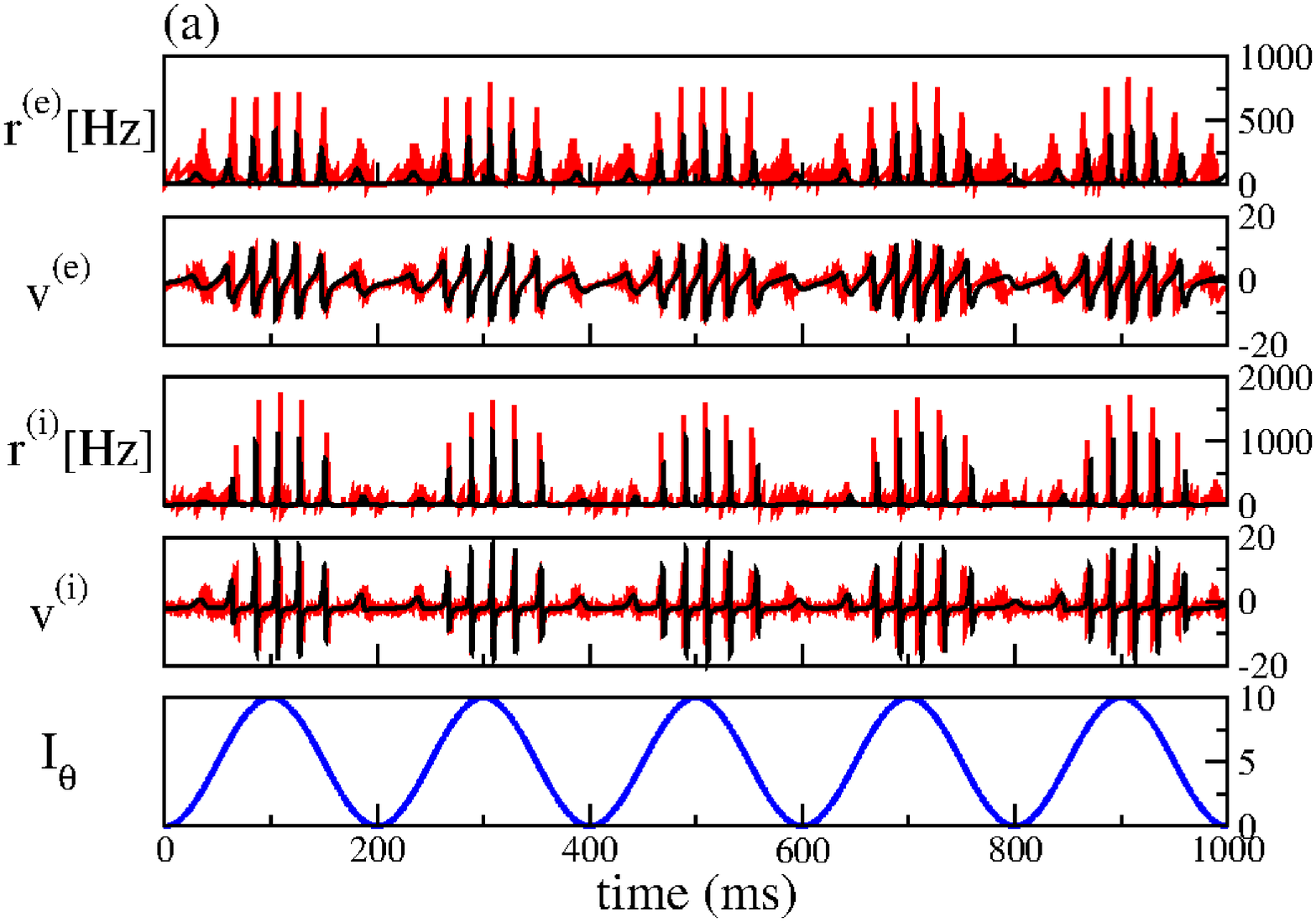}
\includegraphics*[width=0.47\textwidth,clip=true]{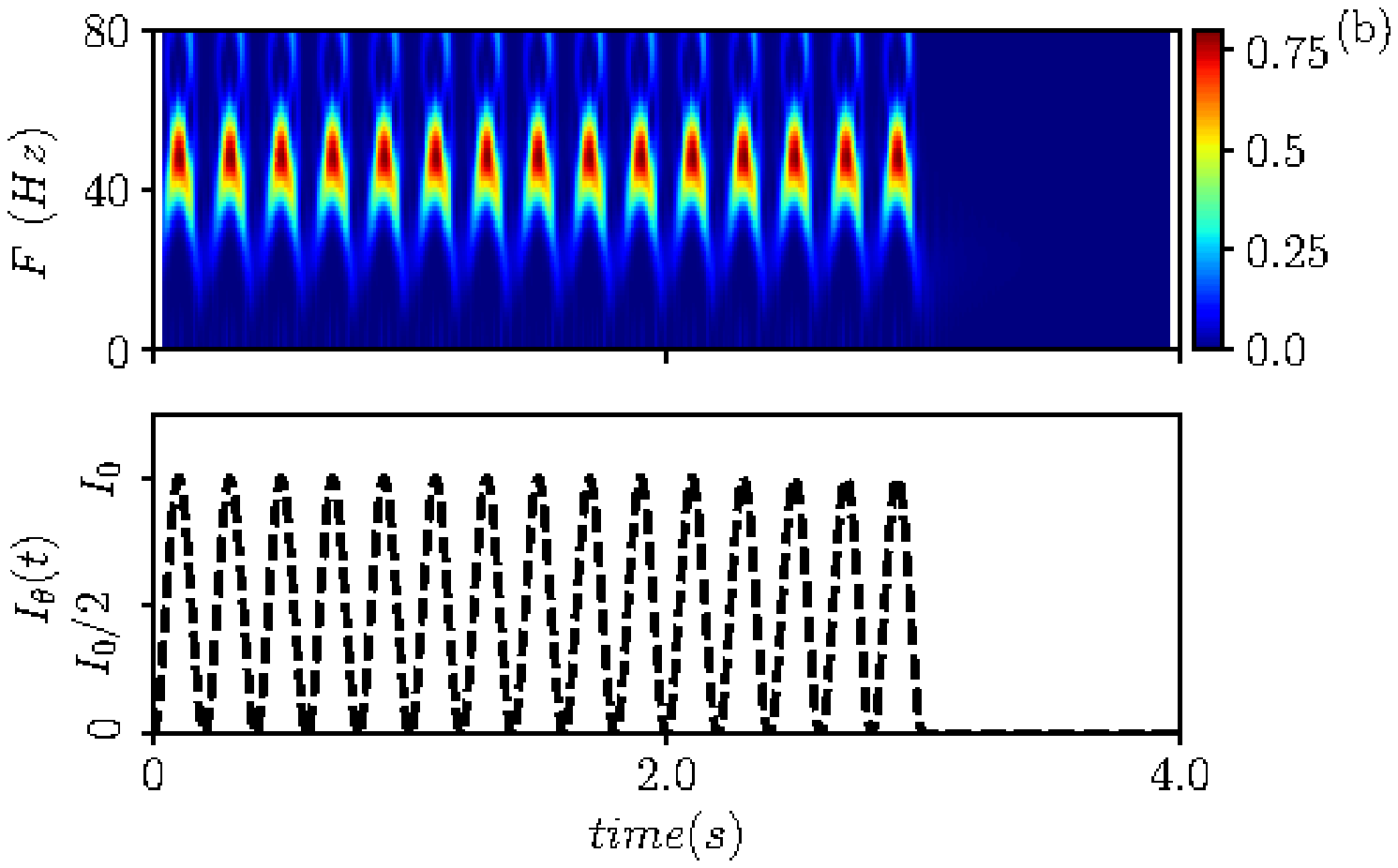}
\includegraphics*[width=0.45\textwidth,clip=true]{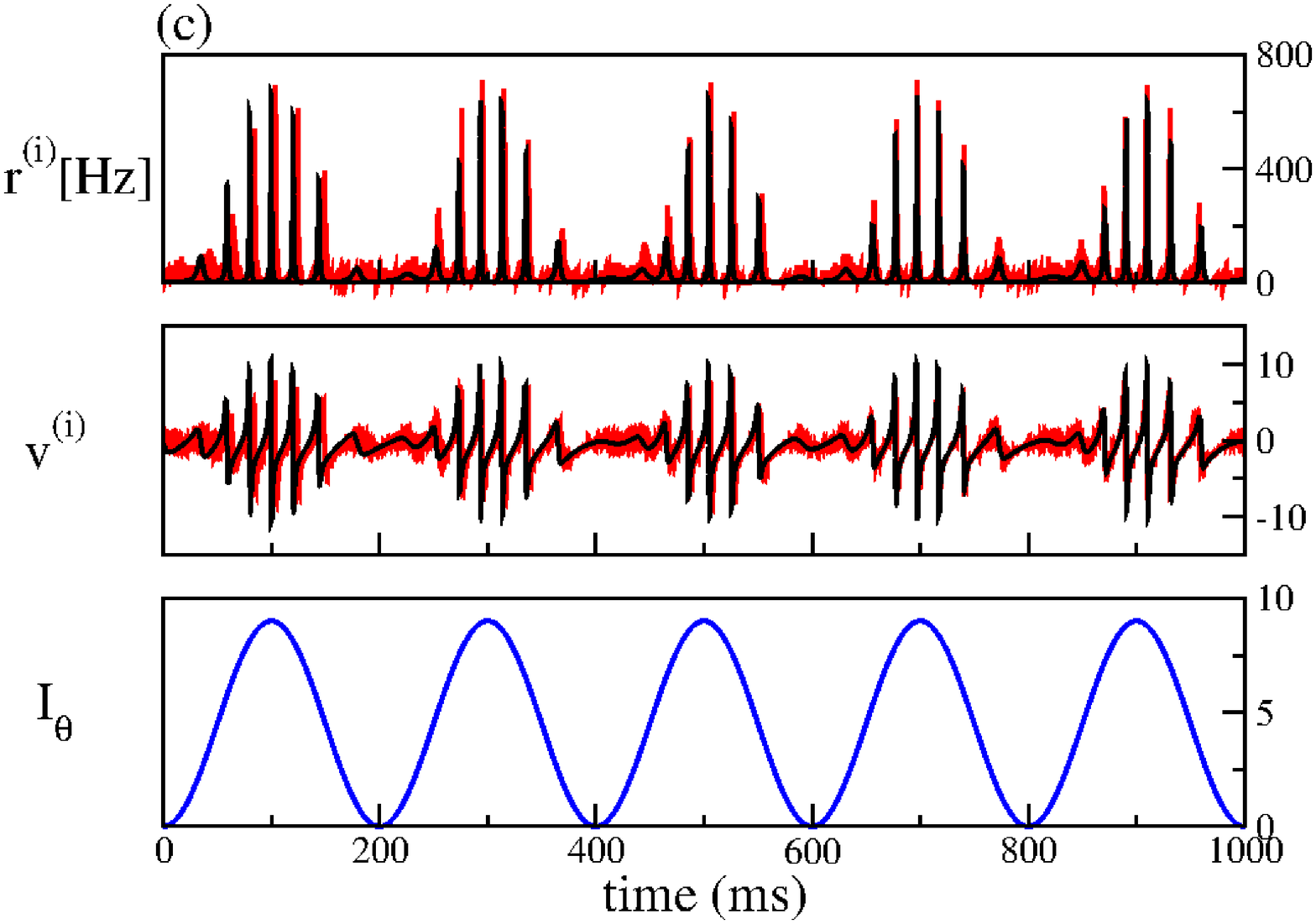}
\includegraphics*[width=0.47\textwidth,clip=true]{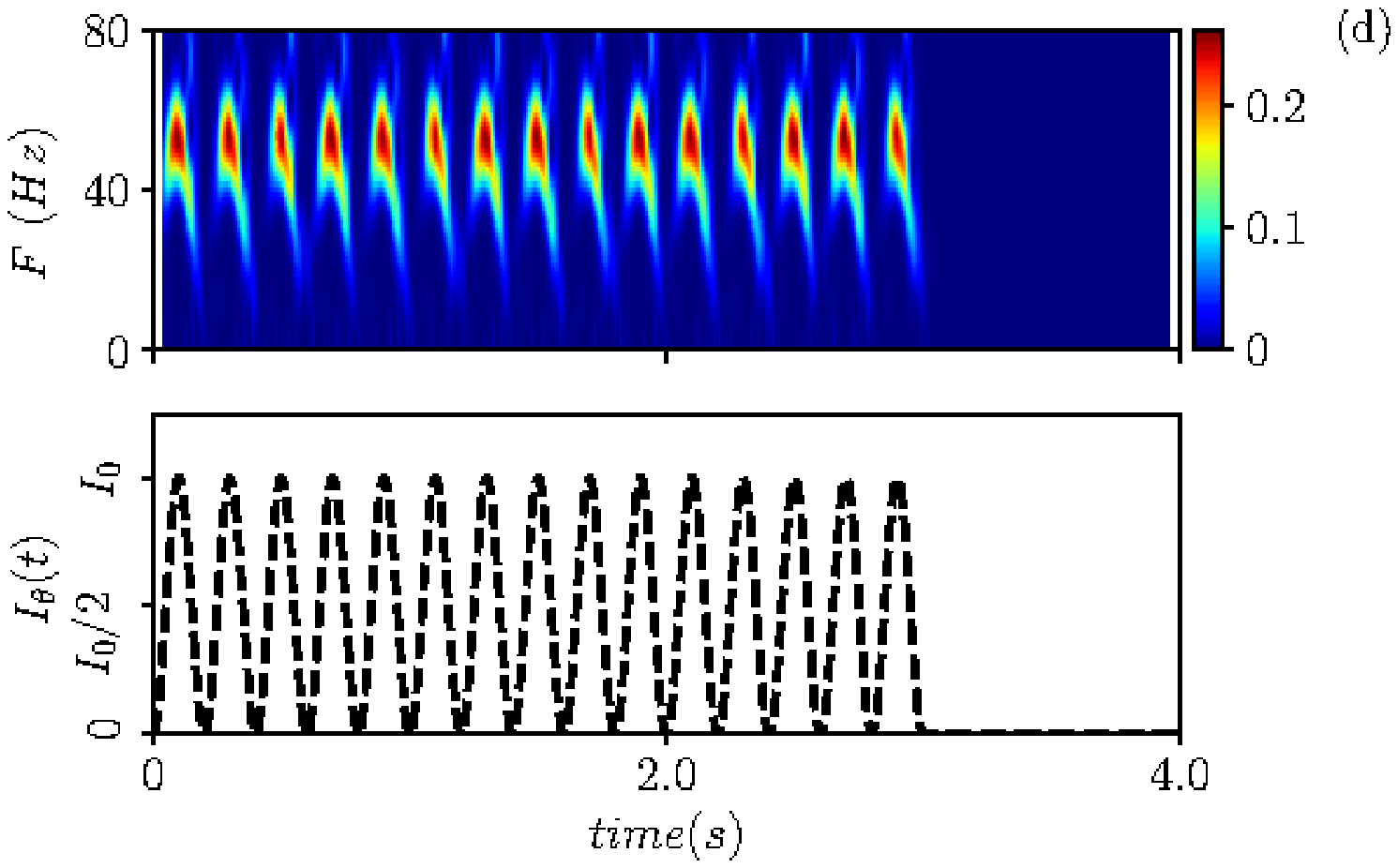}
\end{center}
\caption {{\bf Theta-nested gamma oscillations} ({\bf PING set-up}) (a) From top to bottom: temporal traces of $r^{(e)}$, $v^{(e)}$, $r^{(i)}$, $v^{(i)}$, for the spiking network (red curves) and the neural mass model (black curves). $I_{\theta}$, reported in the bottom panel in blue, is the external current \eqref{current}. For the neural mass model the average rates and membrane potentials are the solutions of Eqs. \ref{eq:macroscopic_ping}, while for the network they are calculated according to Eqs. \ref{indicators}.
(b) Spectrogram of the mean membrane potential $v^{(e)}$ (top) as a function of the external forcing (bottom).
The amplitude of the forcing is $I_0=10$ and its frequency is $\nu_\theta = 5$ Hz. Parameters of the system: $J^{(ee)} = 8$, $J^{(ie)}=J^{(ei)} = 10$, $J^{(ii)} = 0$, ${H}_0^{(e)} = 1.3$, ${H}_0^{(i)} = -5.0$, $\Delta^{(e)} = 1$, $\tau_m^{(e)} = 20$, $\Delta^{(i)} = 1$, $\tau_m^{(i)} = 10.0$, $A=0$, network size $N^{(e)} = N^{(i)} = 5000$. The average firing rates are ${\bar R}^{(e)} \simeq 37$ Hz, ${\bar R}^{(i)} \simeq 36$ Hz.
({\bf ING set-up}) (c) From top to bottom: temporal traces of $r^{(i)}$, $v^{(i)}$ where the line colors have the same meaning as in panel (a).
For the neural mass model, average rates and membrane potentials are solutions of Eqs. \ref{eq:macroscopic_inh}.
(d) Spectrogram of the mean membrane potential $v^{(i)}$ (top) as a function of the external forcing (bottom).
The amplitude of the forcing is $I_0=9$ and its frequency is $\nu_\theta = 5$ Hz. Parameters of the system: $J^{(ii)} = 21.0$, ${H}_0^{(i)} = 2.0$, $\Delta^{(i)} = 0.3$, $\tau_m^{(i)} = 10.0$ ms, $\tau_d = 10.0$ ms, $A=0$, system size for the purely inhibitory network $N^{(i)} = 10000$. The corresponding average firing rate is ${\bar R}^{(i)} \simeq 28$ Hz.}
\label{Fig2_Ping}
\end{figure*}

As a first step, we have verified that the reduced mean-field models 
are able to well reproduce the macroscopic evolution of the spiking network
in both considered set-ups, under the external forcing \eqref{current}.
In particular, we set the unforced systems in the asynchronous regime in proximity of a super-critical Hopf bifurcation,
by choosing $H_0^{(e)} = 1.3 < H^{(e)}_{c}$ and $H_0^{(i)}=-5$ ($H_0^{(i)} =2.0 < H^{(i)}_{c}$) and considered a forcing term with
frequency $\nu_\theta = 5$ Hz and amplitude $I_0 = 10$ ($I_0 = 9$) for the PING (ING) set-up. 

The comparisons, reported in Figs. \ref{Fig2_Ping} (a) and (c),
reveal a very good agreement in both set-ups between 
the network and the neural mass simulations, for 
the mean membrane voltages and the instantaneous firing rates.
Furthermore, in both cases, we clearly observe COs, 
whose amplitudes are modulated by the amplitude of $\theta$-forcing term \eqref{current},
suggesting that we are in presence of a Phase-Amplitude Coupling (PAC) mechanism \citep{hyafil2015}.
The corresponding spectrograms shown in Figs. \ref{Fig2_Ping} (b) and (d) reveal that the
frequencies of the COs are in the $\gamma$-range with the maximum power localized around 50-60 Hz.
Moreover the spectrograms indicate that the process is stationary and due to the external stimulation. The gamma oscillations repeat 
during each $\theta$-cycles and they arrest when the external stimulation is stopped.
The characteristics of these COs resemble $\theta$-nested $\gamma$ oscillations reported in many experiments for neural systems {\it in vitro}
under optogenetic stimulation \citep{akam2012,pastoll2013,butler2016,butler2018} as well as
in behaving animals \citep{chrobak1998}.

\subsection{Wavelet Analysis}

\begin{figure*}[ht]
\includegraphics*[width=\linewidth]{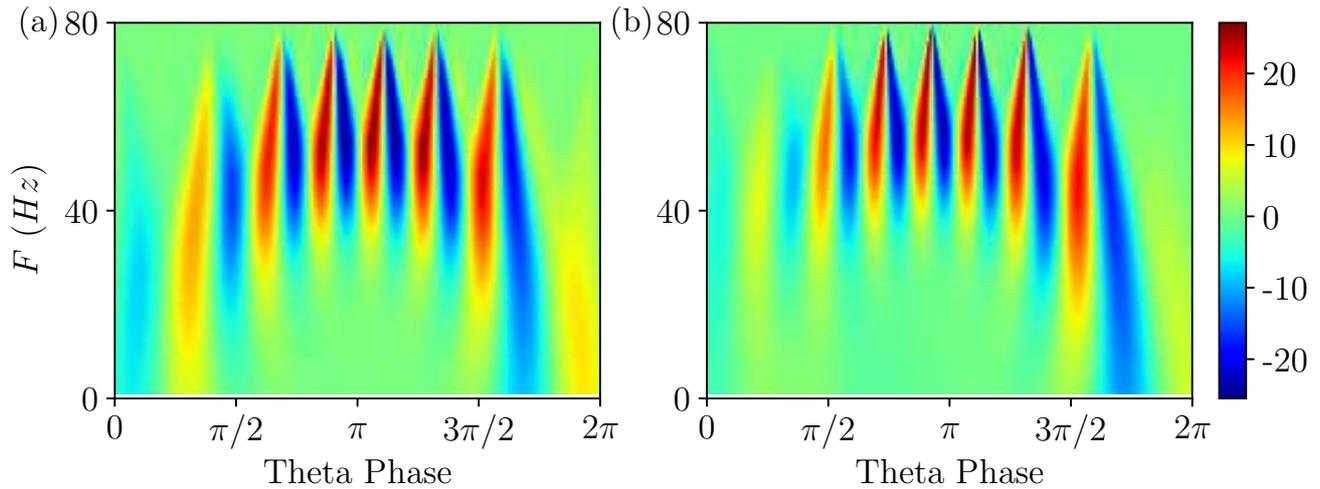}
\caption{{\bf Wavelet Analysis} Continuous wavelet transform over a single $\theta$-cycle 
of the mean membrane potentials $v^{(e)}$ and $v^{(i)}$ appearing in the neural mass models 
for PING (a) and ING (b) set-up, respectively. This analysis allows for accurate automated detection and extraction of gamma activity without the need for bandpass filtering. Parameters as in Fig. \ref{Fig2_Ping}.
}
\label{Fig2wa}
\end{figure*}

To have a deeper insight on these dynamics we have estimated
the continuous wavelet transform of the average membrane potential
on each $\theta$-cycle. As an example, we report in Fig. \ref{Fig2wa} the wavelet spectrogram of the mean potential within a single $\theta$-cycle for 
the previously examined PING (panel (a)) and ING (panel (b)) set-ups. 
Indeed, from the comparison of panel (a) and (b) in Fig. \ref{Fig2wa}, we pratically do not observe any difference: the system responds with COs in the range $[40,80]$ Hz and it exhibits alternating maxima and minima in the wavelet spectrogram as a function of the $\theta$-phase. Similar results have been
reported in \citep{butler2016} for the CA1 region of rat Hippocampus under optogenetic sinusoidal $\theta$ stimulation. 

Differences among the two cases appear when one considers the wavelet spectrograms averaged over many $\theta$ periods: for the PING case the spectrogram remains unchanged, 
instead for the ING set-up the spectrogram smears out and it does not present anymore the clear oscillations reported in Fig. \ref{Fig2wa} (b). 
This difference indicates that in the PING case the observed pattern repeats exactly over each cycle: 
$\gamma$ and $\theta$ oscillations are perfectly phase locked.
This is not the case for the ING set-up: despite the PAC patterns appear quite similar in successive cycles, as shown in Fig. \ref{Fig2_Ping} (c),
indeed they do not repeat exactly. From a point of view of nonlinear dynamics, the PING case
would correspond to a perfectly periodic case, while the other case could be
quasi-periodic or even chaotic. Therefore, we can observe PAC with associated phase locking or in absence of locking.

Furthermore, according to the data shown in Fig. \ref{Fig2wa}, this can also represent an example of PFC, since COs with frequencies $\simeq 40$ Hz occur 
at small and large $\theta$-phases, while in the middle range $\pi/2 < \theta < 3 \pi/2$ one observes similar oscillations with $F \simeq 60$ Hz.
 
\subsection{Phase-Amplitude Locked and Unlocked States}

\begin{figure*}[ht]
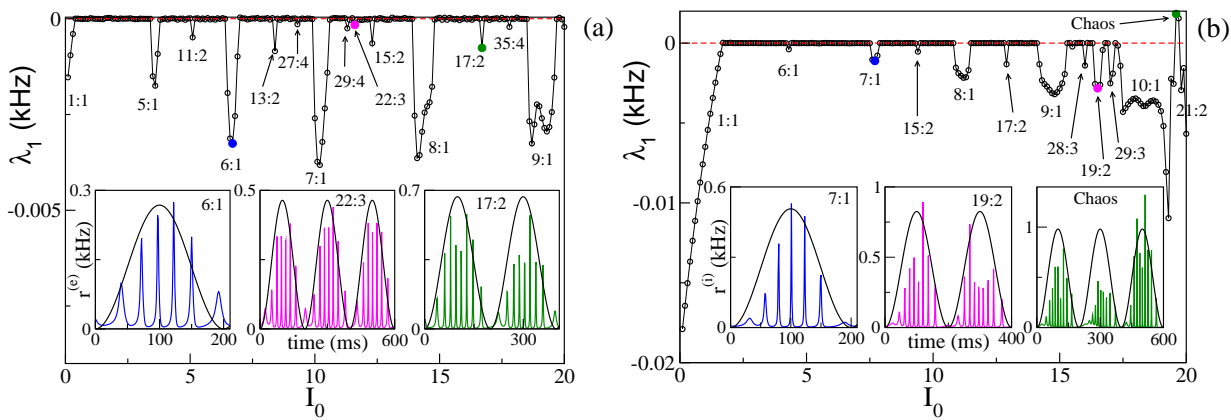

\includegraphics*[width=80mm,clip=true]{fig6a.eps}
\includegraphics*[width=80mm,clip=true]{fig6b.eps}
\caption{{\bf Maximal Lyapunov exponent} $\lambda_1$ estimated for the neural mass model as a function of the forcing amplitude $I_0$,
for the PING (a) and ING (b) set-ups. In bothe cases the system is subject to a forcing frequency $\nu_\theta=5$ Hz.
Insets in panel a (b) report the istantaneous firing rate $r^{(e)}(t)$ ($r^{(i)}(t)$ ) versus time for the PING (ING) set-up respectively. 
The shown three cases are representative of the states identified by circles in the main panels. The color code is the same, i.e. the color used in the inset identifies
the corresponding circle in the main panel. The black continuous lines in the inset corresponde to $I_\theta$ in arbitrary units.
Parameters as in Fig. \ref{Fig2_Ping}.
}
\label{Lyap}
\end{figure*}

To better examine the dynamical regimes emerging in our set-ups we have firstly estimated the maximal Lyapunov exponent $\lambda_1$
associated to the neural mass models, for the same parameters considered in Fig. \ref{Fig2_Ping},  
over a wide range of forcing amplitudes, namely $0 \le I_0 \le 20$.
From the results reported in Figs. \ref{Lyap} (a) and (b), it is clear that $\lambda_1$ is always zero, 
apart from some limited intervals where it is negative. This means that the dynamics is usually
quasi-periodic, apart from some Arnold tongues where there is perfect locking between the external forcing and the forced system. 

We notice that for small amplitudes the forcing entrains the system in a $1:1$ periodic locking, therefore the istantaneous firing
rate displays one peak for each $\theta$-period with the same frequency as the forcing $\nu_\theta$.
This locking is present in a wider region in the ING case (namely, $I_0 < 1.70$) with respect to the PING set-up (namely, $I_0 < 0.40$).
More interesting locking regimes, where the forced populations oscillate in the $\gamma$ range, emerge at larger $I_0$. 
These locking regimes can be considered as $\theta$-nested $\gamma$ oscillations; mostly of them are of the type $m:1$, with $m \in [5:10]$, 
which means that, for each $\theta$-period, the firing rate of the forced populations has $m$ maxima (for specific examples see the insets of Fig. \ref{Lyap} (a) and (b)).
In extremely narrow parameter intervals other, more complex, kinds of locking emerge
of the type $m:n$, where exactly $m$ maxima in the population activity appear for every $n$ $\theta$-oscillations. 
In the examined cases we have identified locked patterns with $n$ up to four. Moreover, for the ING case, we have observed even a chaotic region (see Fig. \ref{Lyap} (b)),
which emerges at quite large forcing amplitude $I_0 \simeq 19$. On the basis of our analysis we
cannot exclude that chaos could emerge also in the PING set-up, for sufficiently strong forcing.

\begin{figure*}[ht]
\begin{center}
\includegraphics*[width=\linewidth]{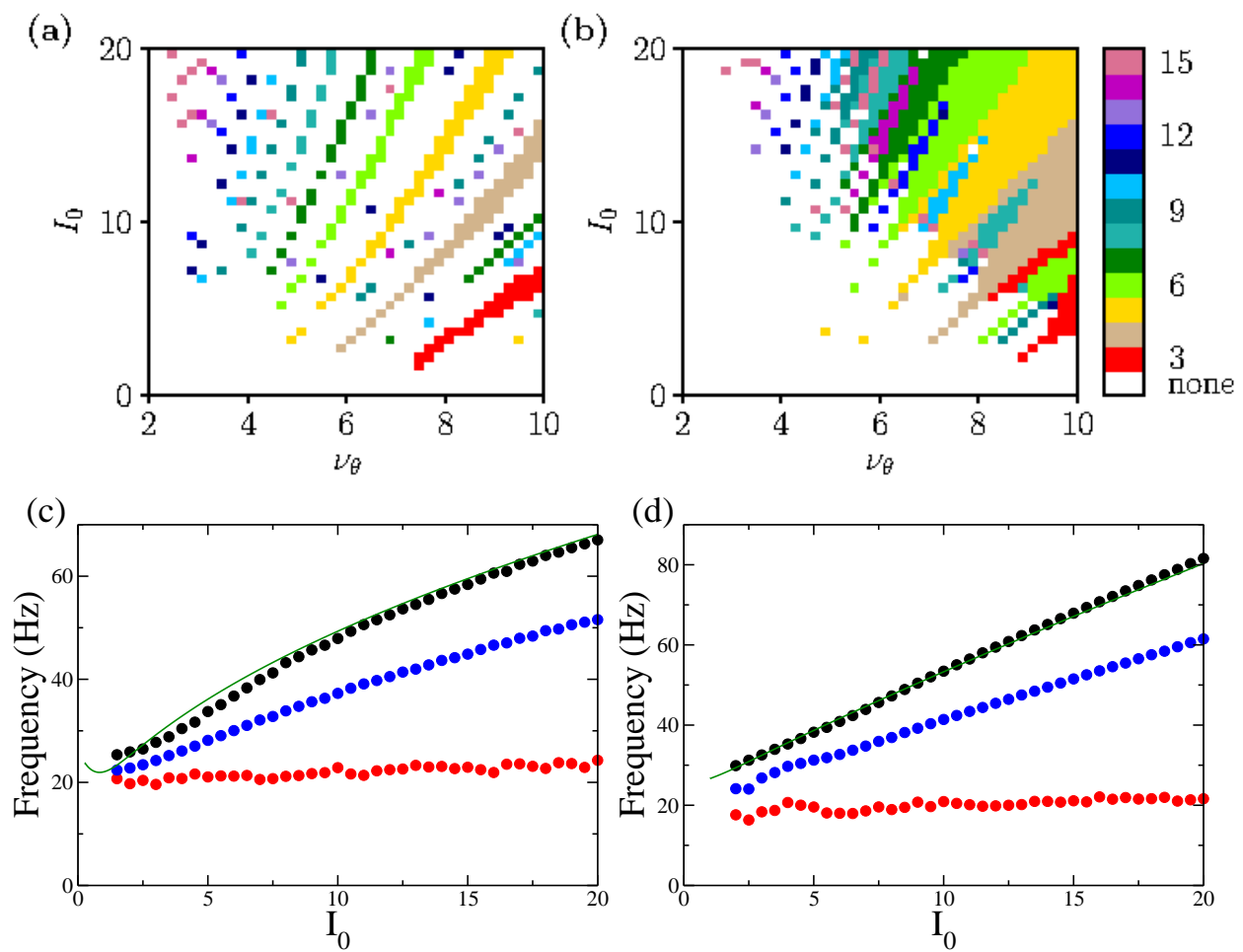}
\includegraphics*[width=0.45\linewidth]{fig7b.eps}
\includegraphics*[width=0.45\linewidth]{fig7c.eps}
\end{center}
\caption{{\bf Phase locked $m:1$ states} Locked states are displayed in panels (a) and (b) 
for the PING and ING set-ups, respectively. 
The color code identifies the locked states accordingly to $m$, from 3 to 15.
(c,d) Minimal (red circles), average (blue circles) and maximal (black circles) frequencies of the COs as a function of the forcing amplitude $I_0$
for PING (c) and ING (d) set-ups. These values are obtained by considering all possible $m:1$ locked states
corresponding to the examined $I_0$. The frequencies $\nu^{(e)}$ ($\nu^{(i)}$) (green solid lines) of the COs obtained from the bifurcation analysis 
in the adiabatic set-up are reported as a function of $H^{(e)} - H_0^{(e)}$ ($H^{(i)} - H_0^{(i)}$)
for the PING (ING).  Parameters as in Fig. \ref{Fig2_Ping}.}
\label{Fig2}
\end{figure*}

Let us now focus on the $m:1$ perfectly locked states with $m>1$, which are worth investigating due to their relevance
for $\theta$-$\gamma$ mixed oscillations as well as to their relative large frequency of occurence with respect to more complex $m:n$ locked states. In particular,
we have examined the response of the system to different forcing amplitudes $I_0 \in [0:20]$
and frequencies $\nu_\theta \in [1:10]$ Hz.  The $m:1$ locked oscillations are reported in Figs. \ref{Fig2} (a) and (b) and 
characterized by the number $m$ of oscillations displayed within a single $\theta$-cycle. 

These locked states appear only for $\nu_\theta > 2-3$ Hz. Moreover the states with equal $m$ are
arranged in stripes in the $(\nu_\theta,I_0)$-plane. Locked states in the PING configuration
occur in separated stripes whose order $m$ increases for increasing $I_0$; in particular
states with $3 \le m \le 10$ are clearly identifiable. In the ING set-up, for sufficiently large $\nu_\theta$ and $I_0$, we have a continuum of locked states, 
thus indicating that, for the ING set-up, phase locking to the forcing frequency is easier to achieve. In this case the order of occurrence of $m$-order states 
is not clearly related to the forcing amplitude; however locked states with order $m$ and $2 m$ are often nested within each other as shown in Fig. \ref{Fig2} (b).

To examine which frequencies are excited in these states we have measured, for each amplitude $I_0$, the minimal,  
the maximal and the average frequency of the COs associated to $m:1$ locked states over the whole range of 
examined forcing frequencies $\nu_\theta$.
These frequencies are reported in Figs. \ref{Fig2} (c) and  (d). The analysis clearly reveals that the minimal CO frequency is essentially
independent from $I_0$ and  its value is around 20 Hz, while the maximal and the average ones grow with $I_0$. However all the frequencies stay 
within the $\gamma$-range for the examined forcing amplitudes. 

To better understand the mechanism underlying the emergence of $\theta$-nested $\gamma$-oscillations,
we have reported in Figs. \ref{Fig2} (c) and (d) the COs' frequencies
$\nu^{(e)}$ ($\nu^{(i)}$) (green solid lines) obtained from the adiabatic bifurcation analysis of the neural mass models
(these frequencies are also shown in the insets of Figs. \ref{Fig3} (a) and \ref{Fig4} (a)).
The  very good agreement between $\nu^{(e)}$ and $\nu^{(i)}$ and the maximal frequency measured for
the locked states suggests that the nested COs are due to the crossing of the super-critical Hopf bifurcation during the periodic stimulation.
In particular, during forcing, the maximal achievable $\gamma$-frequency is the one corresponding to the maximal stimulation current  $I_0 + H_0^{(e)}$ 
($I_0 + H_0^{(i)}$ ) for the unforced PING (ING) set-ups. Furthermore, under sinusoidal forcing, the system spends a longer time in proximity
of the maximal stimulation value, since it is a turning point. This explains why this frequency is always present in the
response of the driven system for the considered locked states.

\subsection{Comparison with Experimental Findings}

In a series of recent optogenetic experiments on the mouse enthorinal-hippocampal
system, have been reported clear evidences that phase-amplitude coupled $\gamma$ rhythms can be generated locally in 
brain slices {\it ex vivo} in the CA1 region, as well as in the CA3 and MEC under 
sinusoidal $\theta$ stimulations \citep{akam2012, pastoll2013, butler2016, butler2018}.
In particular, in \citep{butler2018} the authors reported evidences that,
under theta-rhythmic activation of pyramidal neurons, the generation of the $\gamma$ rhythms is
due to a PING mechanism in all the three mentioned regions. However, due to the
fact that pyramidal neurons are directly activated during experiments, their
result cannot exclude that tonic activation of interneurons contributes to $\theta$–$\gamma$
oscillations {\it in vivo}. Furthermore, in \citep{pastoll2013} the authors affirm that
in MEC $\theta$-nested $\gamma$ oscillations due to the optogenetic $\theta$ frequency drive,
are generated by local feedback inhibition without recurrent excitation, therefore
by a ING mechanism. In this Section we would try to reproduce some of the analyses reported in 
these experimental studies by employing both the PING and ING set-ups, in
order to understand if these two set-ups give rise to different dynamical behaviours.
  
\begin{figure*}[ht]
\begin{center}
\includegraphics*[angle=0,width=80mm,height=70mm,clip=true]{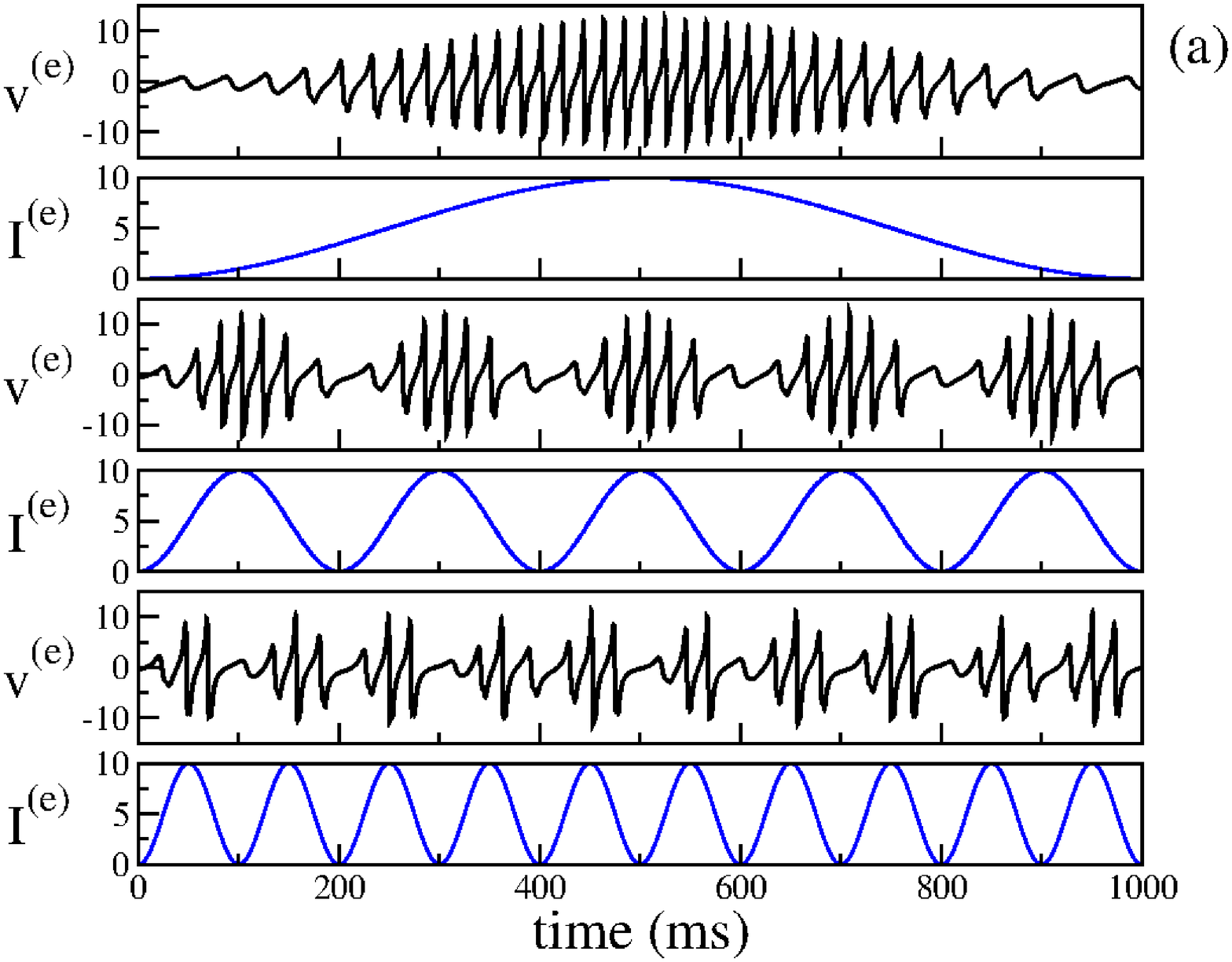}
\includegraphics*[angle=0,width=80mm,height=70mm,clip=true]{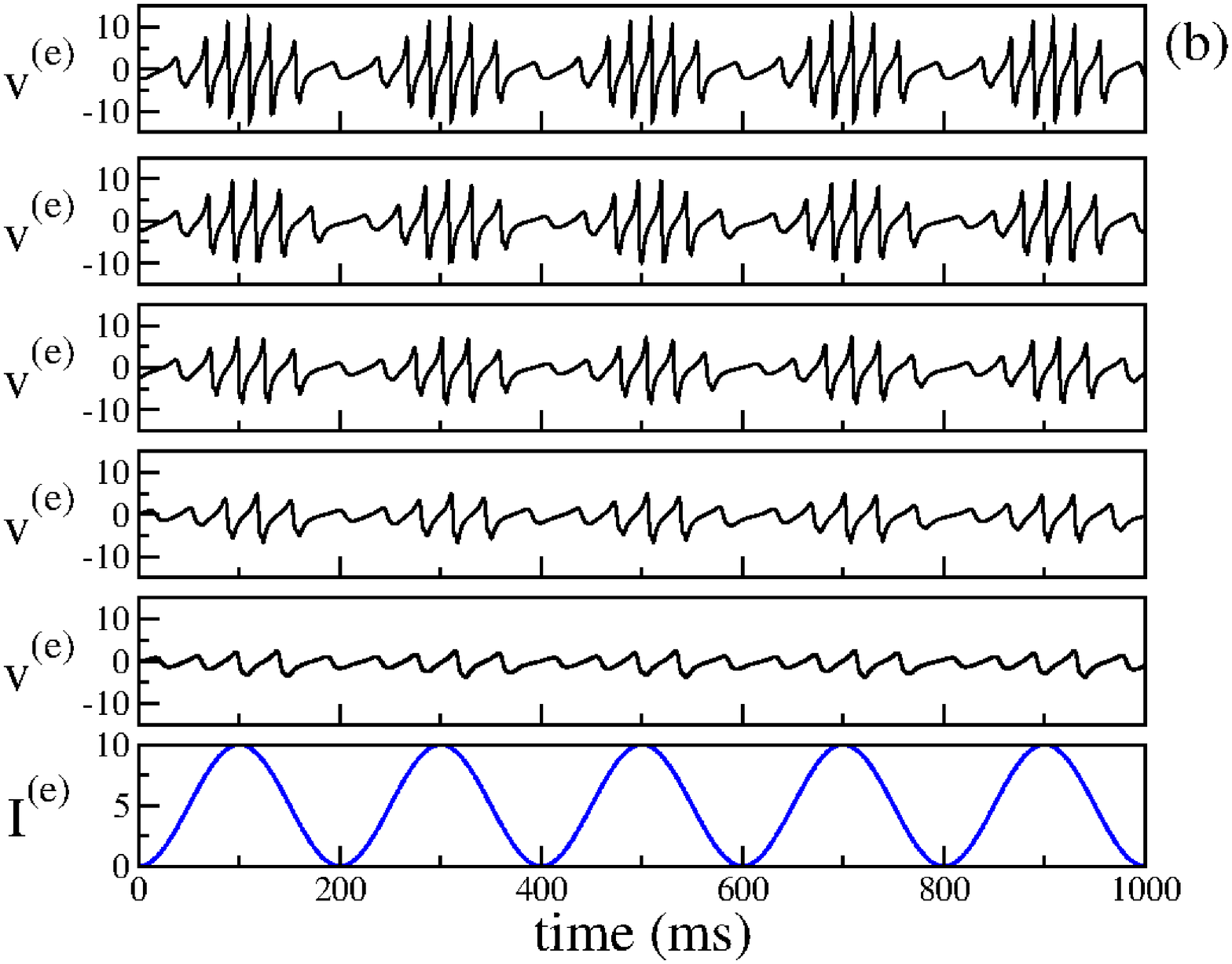}
\includegraphics*[angle=0,width=80mm,height=70mm,clip=true]{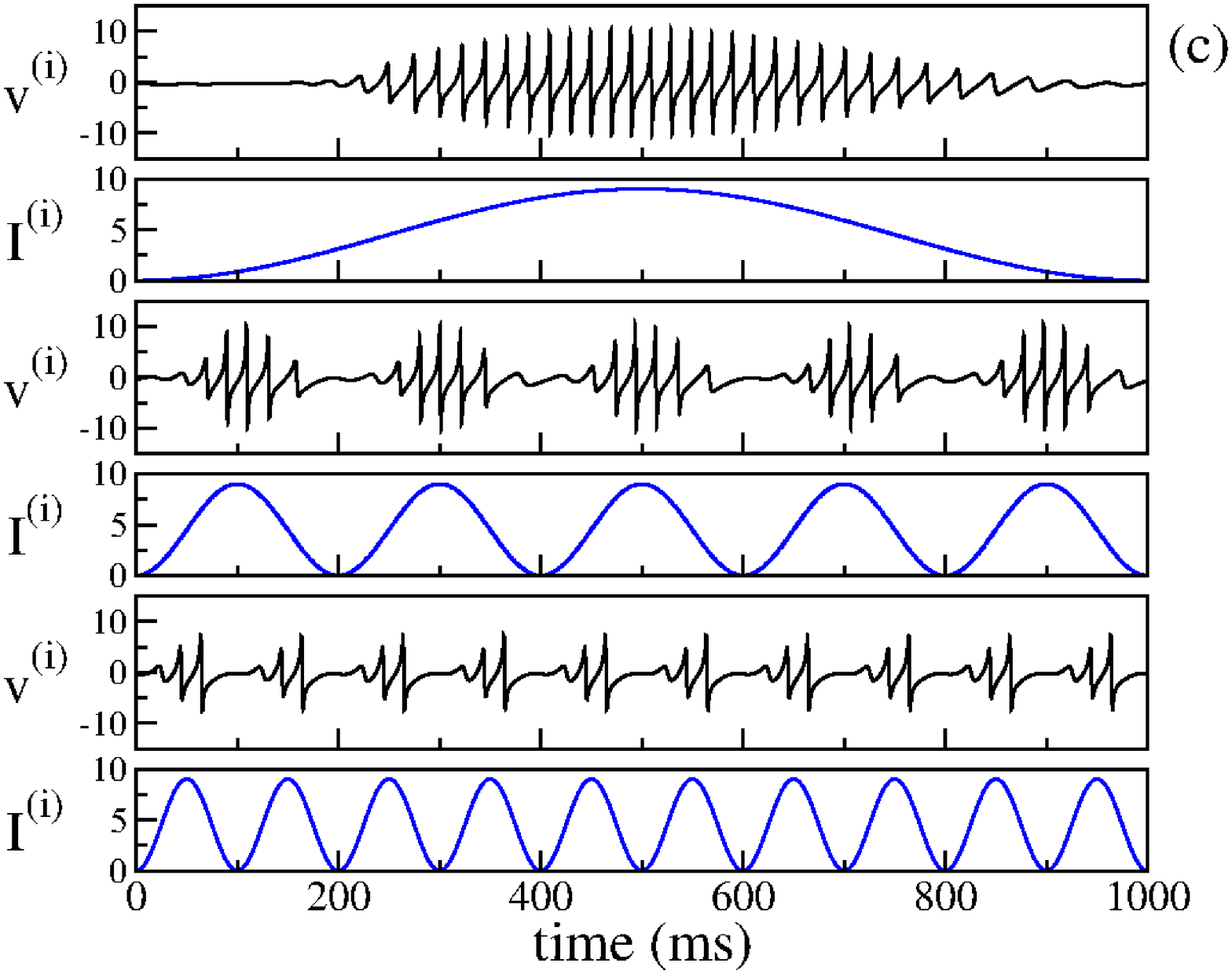}
\includegraphics*[angle=0,width=80mm,height=70mm,clip=true]{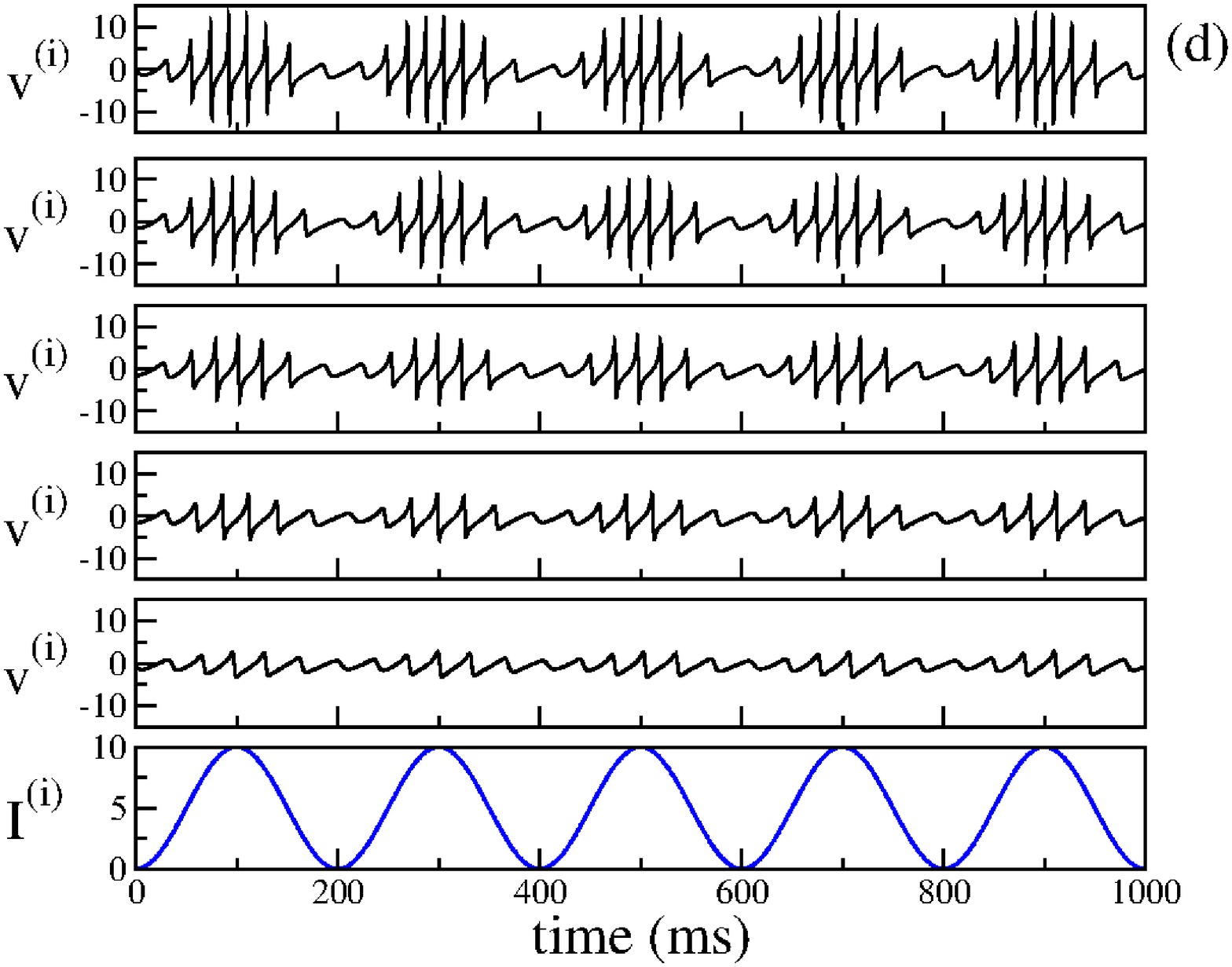}
\end{center}
\caption{{\bf Theta-Nested Gamma COs for PING (a-b) and ING set-up (c-d)}
Left column: dependence of the mean membrane potential of the excitatory (inhibitory) population $v^{(e)}$ 
($v^{(i)}$) on the frequency $\nu_\theta$ of the external forcing $I^{(e)}=I_{\theta}$ ($I^{(i)}=I_{\theta}$)
with $I_0=10$ ($I_0=9$) for the PING (ING) set-up. The current profiles (blue lines) are displayed immediately below 
the corresponding membrane potential evolution. From top to bottom, the frequency $\nu_{\theta}$ is 1 Hz, 5 Hz and 10 Hz.
Right column: dependence of the mean membrane potential $v^{(e)}$ ($v^{(i)}$) 
on the amplitude $I_0$ of the external current. Here the forcing frequency is kept constant at the value $\nu_{\theta} = 5 Hz$.
The amplitude is changed from 20$\%$ of maximum (bottom) to 100$\%$ of maximum (top) in 20$\%$ increments, the maximum being given by $I_0=10$. 
Other parameters as in Fig. \ref{Fig2_Ping}.
}
\label{Fig5}
\end{figure*}

By following the analysis performed in \citep{butler2016,butler2018}, we considered the
response of the two set-ups to forcing of different frequencies $\nu_\theta$
and amplitudes $I_0$. The results reported in Fig. \ref{Fig5} reveal that the phenomenon of PAC is present for all the considered
frequencies $\nu_\theta \in [1,10]$ Hz and amplitudes $I_0 \in [1,20]$ in both set-ups.
Moreover, analogously to what reported in \citep{butler2016,butler2018}, the amplitude of the $\gamma$ oscillations
increases proportionally to $I_0$ while the number of nested oscillations in each cycle
increases for decreasing $\nu_\theta$. On the basis of this comparison, the forced PING and ING
set-ups display essentially the same dynamics.

To get a more detailed information about the dynamics in the two set-ups,
we will now consider the features of the power spectra $P_S^{(e)}$ 
($P_S^{(i)}$) of the mean excitatory (inhibitory) potential
for the PING (ING) set-up. These features are obtained for different forcing amplitudes and
frequencies, somehow similarly to the analysis performed for the power
spectra of the Local Field Potential (LFP) in \citep{butler2016,butler2018}.

\begin{figure*}[ht]
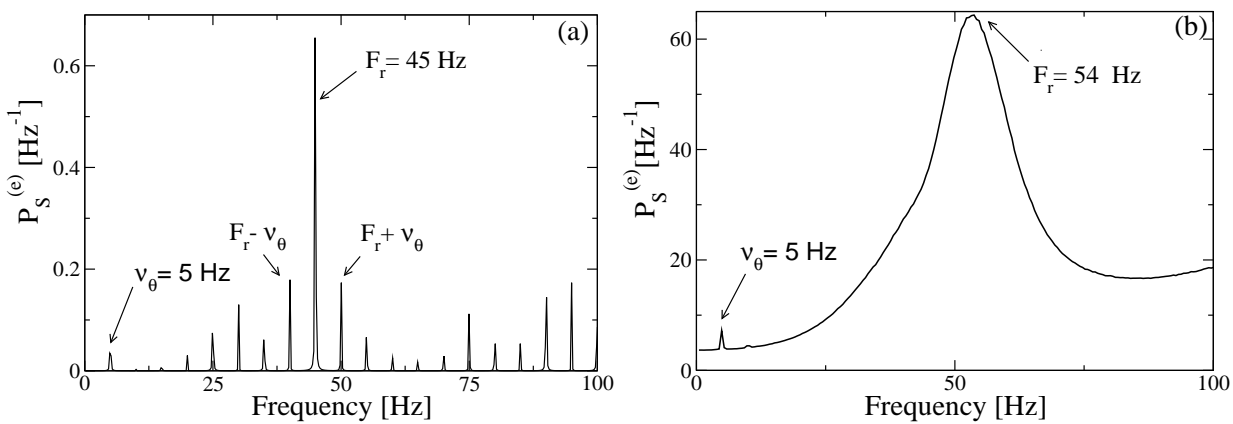

\includegraphics*[width=80mm,clip=true]{fig9a.eps}
\includegraphics*[width=80mm,clip=true]{fig9b.eps}
\caption{{\bf Power spectra for the PING set-up} Spectra $P_S^{(e)}$ of the mean membrane potential $v^{(e)}$ estimated when the excitatory 
population is subject to an external drive with frequency $\nu_{\theta}$ = 5 Hz and amplitude  $I_0=10$, in
absence of noise (a) and for additive noise with amplitude $A=1.4$ (b).
Other parameters as in Fig. \ref{Fig2_Ping}.
}
\label{Fig7}
\end{figure*}

Let us first consider, as an example of the obtained power spectra, the case corresponding to the
PING set-up with a forcing characterized by $\nu_{\theta}$ = 5 Hz and amplitude  $I_0=10$,
shown in Fig. \ref{Fig7} (a). In the spectrum we observe very well defined spectral lines
located at frequencies which can be obtained as a linear combination of the forcing
frequency $\nu_{\theta}= 5$ Hz and of the frequency $F_r=45$ Hz. In particular $F_r$ is associated to the main peak and
should correspond to the intrinsic frequency of the forced system. 
In the present case, the adiabatic bifurcation diagram reported in  Fig. \ref{Fig3} (a) tells
us that  the maximal achievable  COs' frequency is $\nu_{max}^{(e)} \simeq 49.3$ Hz, 
corresponding to $H^{(e)} = I_0 + H_0^{(e)} = 11.3$. Indeed $F_r < \nu_{max}^{(e)}$
and this is due to the fact that the interaction with the forcing system can induce 
a locking phenomenon at a frequency that is exactly a multiple of $\nu_\theta$, as it
happens in the present case. However, in general, a spectrum as the
one shown in  Fig. \ref{Fig7} (a) is the emblem of a quasi-periodic motion characterized
by two uncommensurate frequencies. This can be easily observable in most of the cases in our system,
where $\nu_\theta$ and $F_r$ are usually uncommensurate.

The spectra obtained from optogenetic stimulation, reported in \citep{butler2016,butler2018},
do not resemble the one shown in Fig. \ref{Fig7} (a); indeed they present only two peaks: one
corresponding to the stimulation frequency and one, quite broad, associated to the
$\gamma$ oscillations. We can expect that the difference is due to the multiple noise sources that are
always present in an experimental analyis (and in particular for neurophysiological data), but that are
absent in our model. Indeed, by considering the neural mass model for the PING set-up with
additive noise on the membrane potentials of suitable amplitude, namely $A=1.4$, we get
a power spectrum resembling the experimental ones, as shown in Fig. \ref{Fig7} (b).
The presence of noise induces the merging of the principal peaks in an unique broad one
and the shift of the position of the main peak towards some
larger values (namely, $F_r = 54$ Hz in the present case) with respect to the fully deterministic case.

\begin{figure*}[ht]
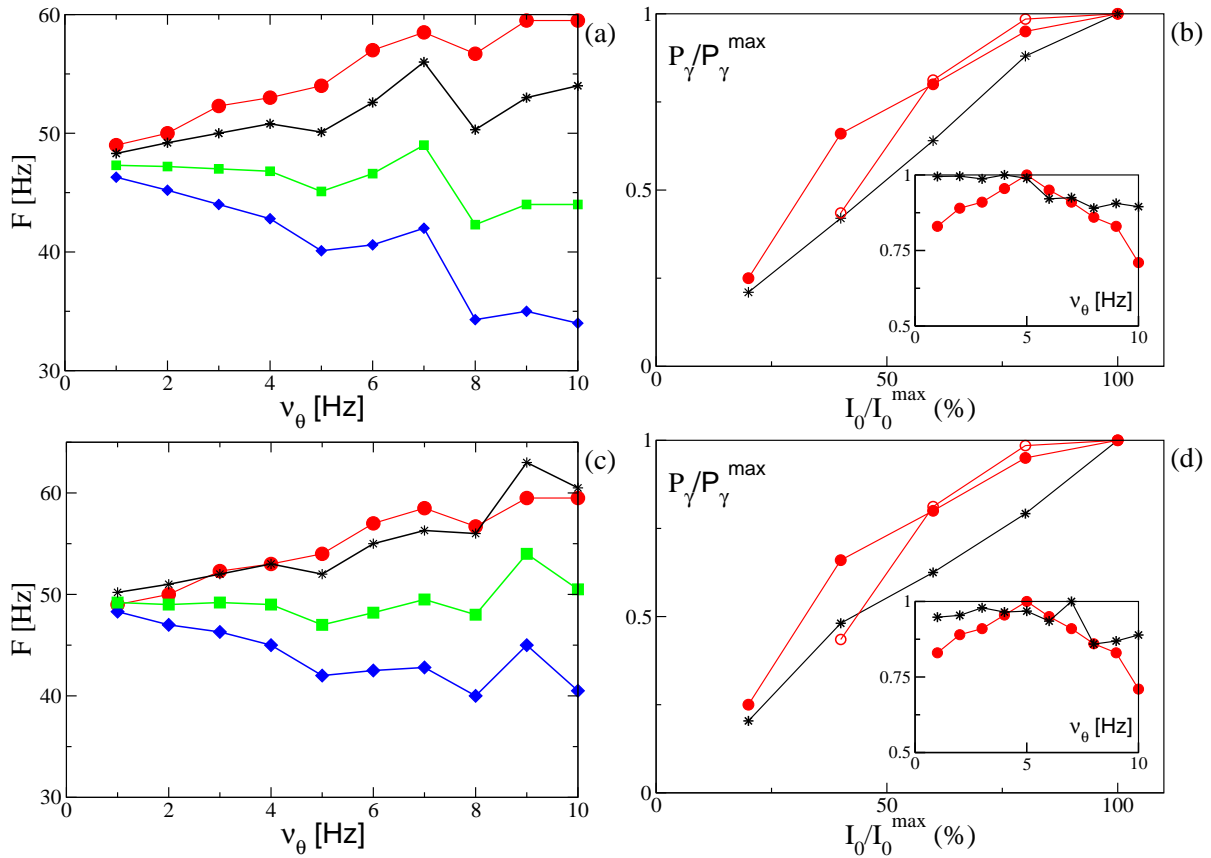

\begin{center}
\includegraphics*[width=0.45\textwidth,clip=true]{fig10a.eps}
\includegraphics*[width=0.43\textwidth,clip=true]{fig10b.eps}
\includegraphics*[width=0.45\textwidth,clip=true]{fig10c.eps}
\includegraphics*[width=0.43\textwidth,clip=true]{fig10d.eps}
\end{center}
\caption{{\bf Power spectra features} ({\bf PING set-up}) (a) Frequencies of the peaks of the power spectrum $P_S^{(e)}$ as a function of the stimulation frequency $\nu_{\theta}$. 
Green squares correspond to the main peak frequency $F_r$, while the black stars to $F_r+\nu_{\theta}$
and the blue diamonds to $F_r-\nu_{\theta}$. The red circles are the experimental data extrapolated from Fig. 4C of \citep{butler2016}. 
The amplitude of the forcing is $I_0=10$. (b) Normalized power of the $\gamma$ oscillations $P_\gamma/P_\gamma^{max}$ associated 
to the signal $v^{(e)}$ as a function of the amplitude stimulation, where we set
$I_0^{max} = 20$ and the frequency of stimulation at $\nu_{\theta}$ = 5 Hz.
In the inset we report the same quantity as a function of the frequency stimulation $\nu_{\theta}$ 
for $I_0=10$. The black stars correspond to our simulations, while the 
red circles to experimental data extrapolated from Fig. 4E (Fig. 4B for the inset) of Ref. \citep{butler2016} (filled circles)
and from Fig. 4C of Ref. \citep{butler2018} (empty circles).
The other parameters are as in Fig. \ref{Fig2_Ping}.
({\bf ING set-up}) (c) Same as in panel (a) for the power spectrum $P_S^{(i)}$ with  $I_0=9$.
(d) Same as panel (b) for the signal $v^{(i)}$ with $I_0^{max} = 40$. For the
inset we set $I_0=9$, other parameters as in Fig. \ref{Fig2_Ping}.
}
\label{Fig9}
\end{figure*}

Let us now consider the power spectra obtained for different forcing frequencies
$\nu_\theta \in [1:10]$ Hz in the $\theta$ range, in case of fixed forcing amplitude
and in absence of noise. The position of the main and auxiliary peaks are shown in Fig. \ref{Fig9} (a) (Fig. \ref{Fig9} (c))
for the PING (ING) set-up and compared with the experimental results (red circles) obtained 
for the CA1 region of the hippocampus in \citep{butler2016}. It is clear that, for both set-ups, the position of the main peak $F_r$
(green squares) has a value $\simeq 50$ Hz and it does not show any clear dependence on  $\nu_\theta$.
This is in contrast with the experimental data, which reveals an increase proportional to $\nu_\theta$ from $49$ Hz
to $60$ Hz. The same trend is displayed in our simulation from the subsidiary peak located at $F_r + \nu_\theta$ (black
stars), that somehow obviously increases with $\nu_\theta$. 

Let us now consider the power of the $\gamma$ oscillations $P_\gamma$ as defined in sub-section II B. 
As shown in the insets of Fig. \ref{Fig9} (b) and (d), this quantity remains essentially
constant for low frequencies (namely, for $\nu_\theta \le 5$ Hz in the PING and
for $\nu_\theta \le 7$ Hz in the ING), while it drops to smaller values at larger frequencies.
On the other hand, the experimental results (red circles) reveal a similar decrease at frequencies
$\nu_\theta > 5$ Hz, but they also reveal an increase at low frequencies, not present
in our numerical data, thus suggesting a sort of resonance at 5 Hz.
For what concerns the dependence of $P_\gamma$ on the forcing amplitude, we have fixed $\nu_\theta = 5$ Hz and varied $I_0$ in
the range $[4:10]$ ($[8:20]$) for the PING (ING) set-up. In both cases and analogously to experimental data, $P_\gamma$ increases
proportionaly to $I_0$, see Fig. \ref{Fig9} (b)  and (d).  

\begin{figure*}[ht]
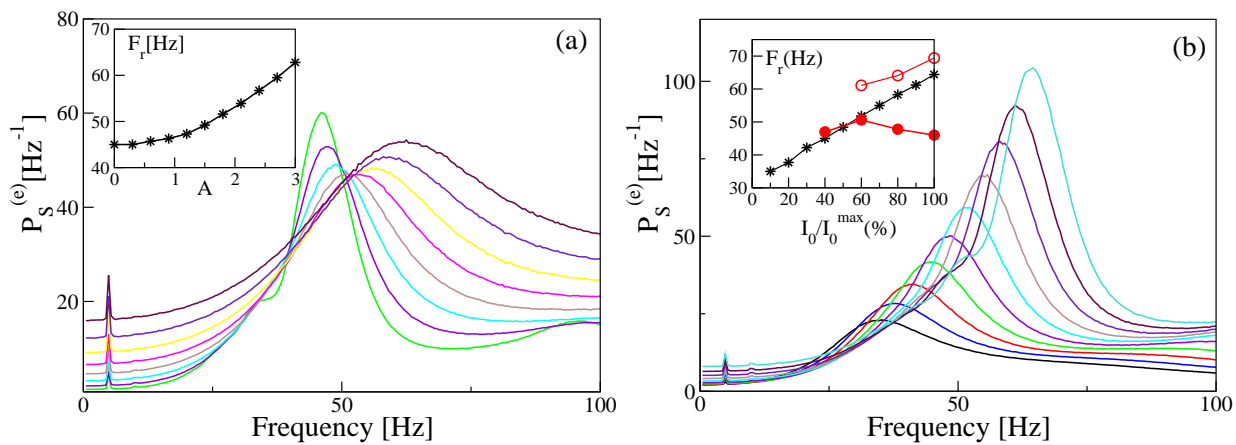

\includegraphics*[width=80mm,clip=true]{fig11a.eps}
\includegraphics*[width=80mm,clip=true]{fig11b.eps}
\caption{{\bf Power spectra dependency on noise and forcing amplitudes} ({\bf PING set-up}) Power spectra $P_S^{(e)}$ 
for different noise level $A$ (a) and different amplitude of the external input $I_0$ (b), for a fixed forcing frequency $\nu_\theta =5$ Hz.
In the insets are reported the frequencies $F_r$ of the main peak as a function of the noise level (a) 
and of the amplitude of the external drive $I_0$ (b). In the inset of panel (b) are also reported experimental data extracted
from Fig. 4F of Ref \citep{butler2016} (filled red circles) and from Fig. 4D of Ref. \citep{butler2018} 
(open red circles).  The curves in (a) are obtained by varying the noise amplitude $A \in [0.9:3.0]$ with a step of 0.3, while keeping fixed $I_0=10$.
On the other hand the curves in (b) refer to different forcing amplitudes $2 \le I_0 \le 20$, varied in steps of 0.2, with
fixed noise amplitude $A=1.4$. The other parameters are as in Fig. \ref{Fig2_Ping}.
}
\label{Fig8b}
\end{figure*}

Our model  is unable to reproduce, in both set-ups, in absence of noise and for fixed forcing amplitude $I_0$, 
the steady increase of $F_r$ with $\nu_\theta$ reported in the experiments for the CA1 of mice in \citep{butler2016}.
Therefore, in order to cope with this problem, we will now investigate how a similar trend can emerge in our data. 
In particular, in the remaining part of the paper we consider noisy dynamics, to have a better match with experiments where it is unavoidable.
In Fig. \ref{Fig8b} (a) we report, for the PING set-up, the estimated power spectra for different noise levels, 
under constant external sinusoidal forcing. The effect of noise is to render the spectrum
more flat and to shift the position of the peak in the $\gamma$ range towards higher
frequencies. As shown in the inset of Fig. \ref{Fig8b} (a), the frequency $F_r$ 
is almost insensitive to the noise up to amplitudes $A \simeq 1.0$, then it increases steadily
with $A$ from $\simeq 45$ hz to $\simeq 62$ Hz. The effect of varying the forcing amplitude $I_0$,
for constant forcing frequency $\nu_\theta =5$ Hz and noise amplitude $A=1.4$, is shown in
Fig. \ref{Fig8b} (b). In this case the amplitude increase of the forcing leads to more defined peaks
in the $\gamma$ range and to an almost linear increase with $I_0$ of $F_r$, as reported in the inset.
In the same inset we have reported also the results related to two optogenetic experiments for the CA1 region
of the mice hippocampus. In particular the data sets refer to two successive experiments performed by the same
group and reported in \citep{butler2016,butler2018}. While in one experiment (red open circles)
a clear increase of $F_r$ with the forcing amplitude is observable
from $60$ to $70$ Hz,  in the another one (red filled circles) the frequency remains almost constant around $45-50$ Hz 
for a variation of $I_0$ from 40 to 100 \% of the maximal amplitude \citep{butler2018}. From this comparison,
we can affirm that our data catch the correct range of frequencies in both experiments and the dependence
on the forcing amplitude reported in \citep{butler2018}.

From the previous analysis we have understood that, for constant forcing frequency, the $\gamma$-peak shifts towards higher frequencies
by increasing the forcing amplitude or the noise level.

\begin{figure*}[ht]
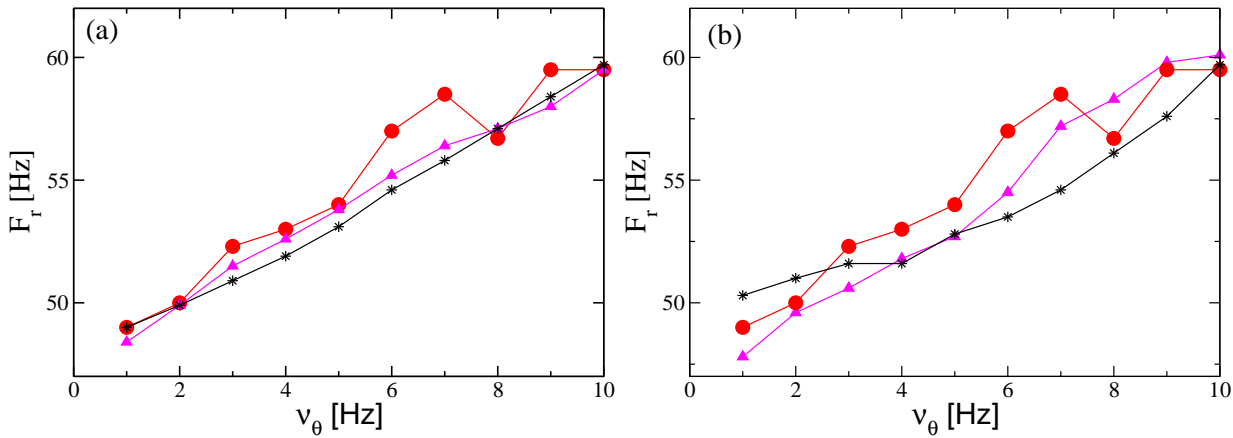

\includegraphics*[width=80mm,clip=true]{fig12a.eps}
\includegraphics*[width=80mm,clip=true]{fig12b.eps}
\caption{{\bf Influence of the theta frequency on the gamma oscillations}
Frequency $F_r$  of the main peak of the power spectrum $P_S^{(e)}$ versus $\nu_{\theta}$ for the PING (a) and ING (b) set-ups. 
Red filled circles represent the experimental data extrapolated from Fig. 4C in \citep{butler2016}.
Black stars (magenta triangles) refer to numerical data obtained by varying linearly the noise amplitude $A$ 
(the forcing amplitude $I_0$) as a function of $\nu_{\theta}$ and maintaining the forcing amplitude $I_0$ (the noise amplitude $A$) constant. 
The data shown as black stars for the PING (ING) set-up in panel (a) (panel (b)) are obtained by adding white noise to the evolution of the 
mean membrane potentials and by varying linearly its amplitude in the interval 
$A \in [1.4 : 2.9]$ as a function of $\nu_\theta$ with $I_0=10$ ($I_0=9$). The magenta triangles refer to data obtained
by keeping fixed the noise amplitude at the value $A=1.4$ and by varying linearly with $\nu_\theta$ the forcing
amplitude $I_0$ in the range $[9.5:18]$ ($[8:14]$) for the PING (ING) set-up in panel (a) (panel (b)).
Other parameters for as in Fig. \ref{Fig2_Ping}.
}
\label{Fig10}
\end{figure*}

Therefore to obtain an increase of $F_r$ with the forcing frequency $\nu_\theta$,
analogously to the results reported in \citep{butler2016} (and displayed as filled red circles in Fig. \ref{Fig9}(a) and (c)), 
we should perform numerical experiments where $\nu_\theta$ increases together with $A$ or $I_0$. The simplest  protocol is to assume that $A$ ($I_0$) will
increase linearly with $\nu_\theta$. The results obtained for the PING (ING) set-up
are reported in Fig. \ref{Fig10} (a) (Fig. \ref{Fig10} (b)). As evident from the figures, in both set-ups
and for both protocols we obtain results in reasonable agreement with the experiment.
In the present framework, we have also analyzed the dependence of the $\gamma$-power $P_\gamma$ on $\nu_\theta$.
In particular this quantity increases almost linearly with the forcing frequency, at variance
with the experimental results in  \citep{butler2016} which revealed a sort of resonance
with an associated maximal $\gamma$-power around $\nu_\theta =5$ Hz (the experimental data 
are displayed as red circles in the insets of Fig. \ref{Fig9}(b) and (d)).

\section{Discussion and Conclusions}

In this paper we have analyzed the dynamics of a new class of neural mass models
arranged in two different set-ups: an excitatory-inhibitory network (or PING
set-up) and a purely inhibitory network (or ING set-up). 
The considered neural mass models are extremely relevant to mimick neural
dynamics for two reasons. On one side because they are not derived heuristically,
since they reproduce exactly the dynamics of excitatory and inhibitory networks of spiking neurons
for any degree of synchronization \citep{montbrio2015,devalle2017,ceni2019}. 
On another side these neural masses reproduce the macroscopic dynamics
of quadratic integrate-and-fire neurons, which are normal forms of class I neurons,
therefore they are expected to represent the dynamics of this large class of neurons
\citep{ermentrout1986}.

In this paper we have shown that $\theta$-nested $\gamma$ oscillations
can emerge both in the PING and ING set-up under an external excitatory $\theta$-drive
whenever the system, in absence of forcing, is in a regime of asynchronous dynamics,
but in proximity of a Hopf bifurcation towards collective $\gamma$ oscillations.
The external forcing drives the system across the bifurcation inside the oscillatory 
regime, thus leading to the emergence of $\gamma$ oscillations.
The amplitude of these collective oscillations is related to the distance from the bifurcation point, therefore it depends
on the phase of the $\theta$-forcing term. These nested oscillations can arise both in proximity of
super-critical and sub-critical Hopf bifurcations. However, in the latter case their amplitudes are no more symmetric with respect to the
maximum value of the theta stimulation, somehow analogously to the experimental
findings reported in \citep{butler2016}.
 
Analogous results have been reported for an excitatory-inhibitory network
with a recurrent coupling among the excitatory neurons, by considering the Wilson-Cowan rate model \citep{onslow2014}. 
However, at variance with our neural mass model, the Wilson-Cowan one fails to reproduce the emergence of $\gamma$-oscillations, 
displayed by the corresponding spiking networks, in several other set-ups.
In particular, the Wilson-Cowan model is unable to display COs for purely inhibitory populations (the ING set-up), without the addition of a delay
in the IPSPs transmission, delay that is not required in the network model.
Moreover  the Wilson-Cowan model is unable to display COs even for excitatory-inhibitory coupled populations
in absence of a recurrent excitation \citep{onslow2014,devalle2017}. As shown in Appendix B, the considered neural mass model in the PING set-up displays
clear $\theta$-nested $\gamma$ oscillations in absence of any recurrent coupling or with recurrent couplings only among the inhibitory neurons.
 
Furthermore, we have identified two different types of phase amplitude couplings.
One characterized by a perfect locking between $\theta$ and $\gamma$ rhythms,
corresponding to an overall periodic behaviour dictated by the slow forcing.
The other one where the locking is imperfect and dynamics is quasi-periodic or even chaotic.
The perfectly locked $\theta$-nested $\gamma$ oscillations display at the same time two types of cross-frequency coupling: 
phase-phase and phase amplitude coupling \citep{hyafil2015}.
These states arise for $\nu_\theta$ larger than 2-3 Hz and for sufficiently large forcing amplitudes.
From the results reported in \citep{butler2016} for the CA1 region of the hippocampus
under sinusoidal forcing {\it in vitro}, it is evident that perfectly phase locked PACs have been observed in each single slice.
However, {\it in vivo} this perfect phase-phase locking cannot be expected, see
the detailed discussion of phase-phase coupling reported in \citep{scheffer2016},
where the authors clarify that phase locking is indeed observable, but only over
a limited number of successive $\theta$-cycles. Therefore, PAC with an underlying chaotic (or noisy) dynamics is the scenario usually expected in behaving animals.

From our analysis it emerges also that locked states are more frequent in the ING set-up.
The purely inhibitory population is more easily entrained by the forcing with respect to
the coupled excitatory-inhibitory population system, where the forcing is applied to the excitatory population.
This result is somehow in agreement with recent findings based on the analysis of
phase response curves, which suggest that stimulating the inhibitory population facilitates the entrainment of the gamma-bands
with an almost resonant frequency \citep{akao2018,dumont2019}. However, these analyses do not 
consider $\theta$-$\gamma$ entrainment: this will be a subject of future studies 
based on exact macroscopic phase response curves \citep{dumont2017,dumont2019}.

Our modelization of the PAC mechanism induced by an external $\theta$-forcing
is able to reproduce several experimentals features reported for optogenetic experiments
concerning the region CA1, CA3 of the hippocampus, as well as MEC
\citep{akam2012,pastoll2013,butler2016,butler2018}. 
In agreement with the experiments, we observe nested $\gamma$ COs for forcing frequencies
in the range $[1:10]$ Hz, whose amplitude grows proportionally to the forcing one. Furthermore, the $\gamma$ power and the frequency of the
$\gamma$-power maximum increase almost linearly with the forcing amplitude.
However, the neural mass model in all the examined PING and ING set-ups is unable to
reproduce the increase in frequency of the $\gamma$-power peak with $\nu_\theta$ reported in \citep{butler2016}. Indeed such effect 
was expected by the observation that during movement, both the frequencies of hippocampal theta oscillations \citep{slawinska1998}
and gamma oscillations \citep{ahmed2012} increase with the running speed of the animal. However, the variation of the 
$\gamma$ frequency reported in \citep{ahmed2012} for behaving animals amounts almost to 100 Hz,
while in the optogenetic experiment by Butler et al. \citep{butler2016}, the increase was limited to $\simeq 10$ Hz. In order to get a similar increase
in the neural mass model, we have been obliged to assume that the noise (or the forcing amplitude)
increases proportionally to $\nu_\theta$. On one side, further experiments are required to clarify 
if, during optogenetic experiments, the forcing (or noise amplitude) affecting the neural dynamics is indeed dependent on $\nu_\theta$. This could be due
to a reinforcement of the synaptic strenghts for increasing forcing frequencies, 
or to the fact that higher $\theta$ frequencies can favour neural discharges in regions
different from CA1, that can be assimilated to external noise.
On another side it should be analysed if other bifurcation mechanisms, beside 
the Hopf one, here considered, can give rise to such a dependence of $\gamma$ power on $\theta$ forcing.
 
\section*{Funding}
AT received financial support by the Excellence Initiative I-Site Paris Seine (Grant No ANR-16-IDEX-008) (together with HB and MS), by the Labex MME-DII (Grant No ANR-11-LBX-0023-01) (together with SO) and by the ANR Project ERMUNDY (Grant No ANR-18-CE37-0014), all part of the French programme ``Investissements d'Avenir''.

\section*{Acknowledgments}
This work has been initiated during 
the Advanced Study Group 2016/17 "From Microscopic to Collective Dynamics in Neural Circuits"
at the Max-Planck Institute for Physics of Complex System in Dresden (Germany), where SO
and AT had extremely useful interactions with E. Montbri\'o and H.P. Robinson.
We also acknowledge fruitful discussions with D. Angulo-Garcia, 
F. Devalle, G. Dumont, M. di Volo, A. Pikovsky.

\section*{Author Contributions}
MS, HB, and SO performed the simulations and data analysis. 
AT was responsible for the state-of-the-art review and the paper write-up.
All the authors conceived and planned the research.

\section*{Conflict of Interest Statement}
The authors declare that the research was conducted in the absence of any commercial or financial relationships that could be construed as a potential conflict of interest.

\appendix

\section*{Appendix A: PING set-up: sub-critical Hopf}

\begin{figure}[ht]
\begin{center}
\includegraphics[width=0.33\textwidth,clip=true]{fig13a.eps}
\includegraphics[width=0.33\textwidth,clip=true]{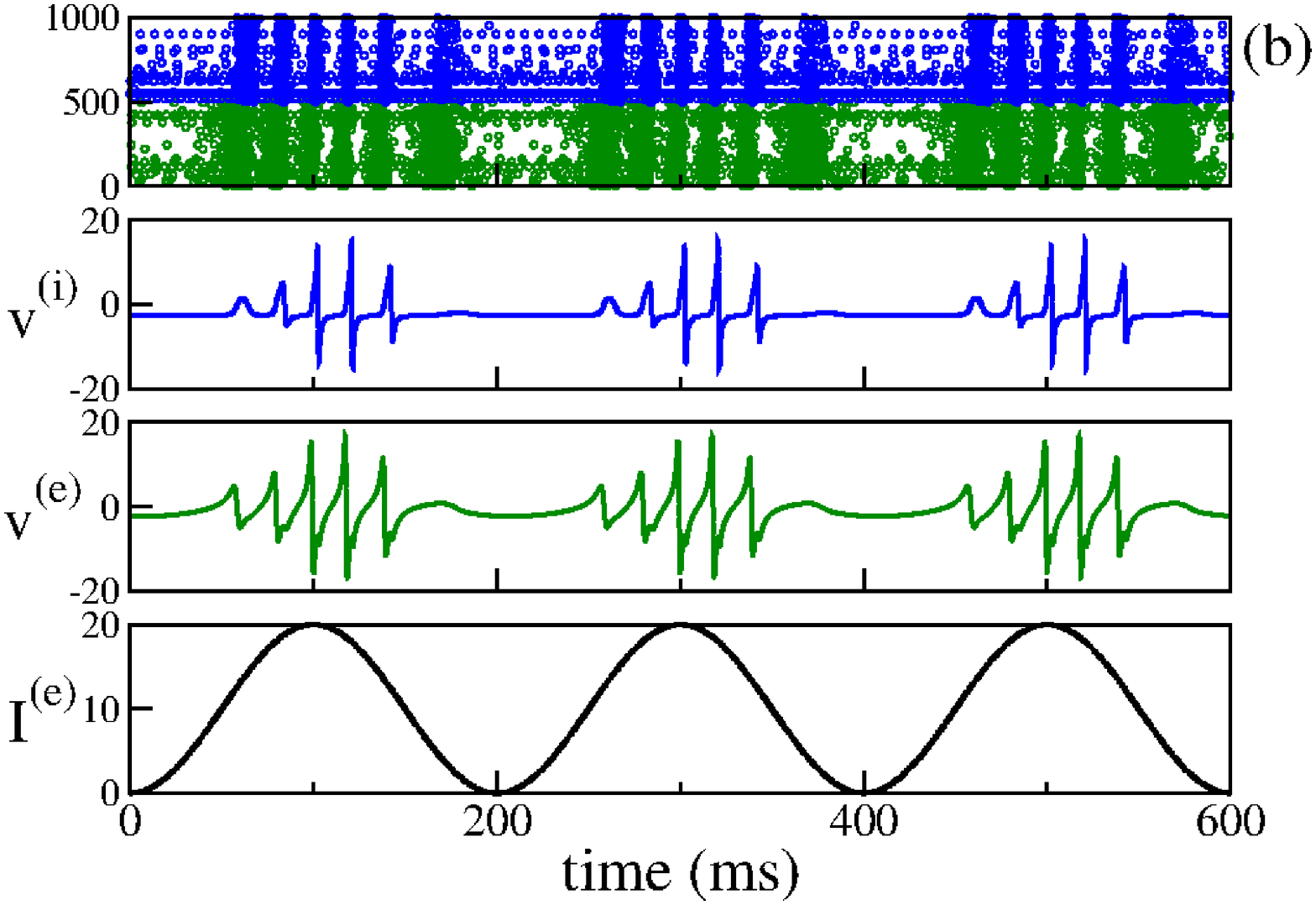}
\includegraphics[width=0.32\textwidth, clip=true]{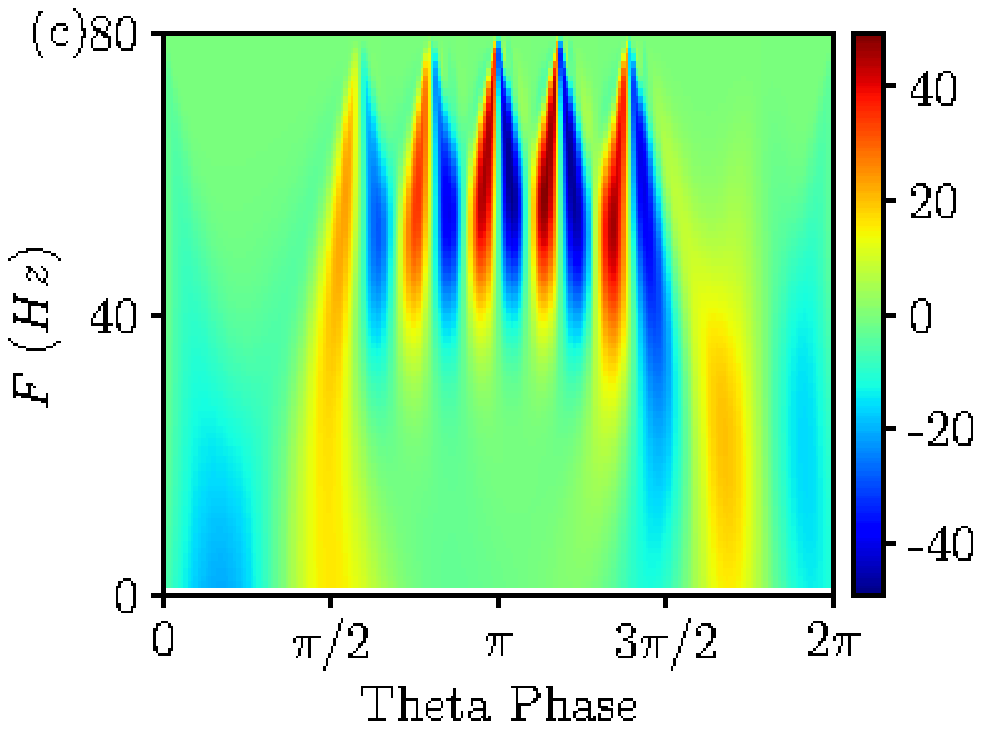}
\end{center}
\caption {({\bf PING set-up: subcritical Hopf}) (a) Bifurcation diagram of the average membrane potential $v^{(e)}$ as a function of $H^{(e)}$.  The black continuous (dashed) line identifies the stable (unstable) fixed point. The red lines denote the extrema of the limit cycles. The subcritical Hopf bifurcation occurs at $H^{(e)}_c \simeq 7.8$ while the saddle-node of limit cycles at $H_{SN}^{(e)} = 5.8$. In the inset the COs' frequency $\nu^{(e)}$ is displayed as a function of $H^{(e)}$. 
(b) From top to bottom: raster plot where green (blue) dots refer to excitatory (inhibitory) neurons;
average membrane potentials $v^{(i)}$ and $v^{(e)}$ as obtained by the evolution
of the neural mass models and forcing current $I^{(e)}$ for $H^{(e)}=-5 < H_{SN}^{(e)}$
and $\nu_\theta = 5$ Hz. (c) Continuous wavelet transform over a single $\theta$ cycles for $v^{(e)}$ with system setting
as in (b). The remaining system parameters are $J^{(ee)}=8$, $J^{(ii)}=0$, $J^{(ie)}=J^{(ei)} = 10$, $H^{(i)}=-8.0$
}
\label{FigPINGsub}
\end{figure}

In the PING set-up with only recurrent excitation (i.e. with $J^{(ee)} \ne 0$, $J^{(ii)} = 0$ and $J^{(ie)} = J^{(ei)} \ne 0$),
it is possible to observe the emergence of COs also via a sub-critical Hopf bifurcation, by using $H^{(e)}$ as control parameter, as shown in Fig. \ref{FigPINGsub} (a).
This is due to the nature of the Hopf bifurcation that can be modified by simply varying the value of $H^{(i)}$.
In this case we observe three regimes: an asynchronous one for $H^{(e)} < H_{SN}^{(e)} $; an
oscillatory one for $H^{(e)} > H_{c}^{(e)} $ and a bistable one in the range $[H_{SN}^{(e)}:H_{c}^{(e)}]$.
The frequency of the COs $\nu^{(e)}$ is always in the $\gamma$ range with a minimal value
$\simeq 36$ Hz achievable at the Hopf bifurcation, see the inset of Fig. \ref{FigPINGsub} (a).

If we consider the unforced system with $H^{(e)} < H_{SN}^{(e)}$ and we apply a $\theta$-forcing, we observe PAC oscillations. 
However when considering $v^{(e)}$, the COs are now asymmetric with respect to the maximum of the stimulation 
current $I^{(e)} = I_\theta(t)$ (see Fig. \ref{FigPINGsub} (b)). 
This effect is even more pronounced by observing the wavelet spectrogram reported in
Fig. \ref{FigPINGsub} (c), where a clear PFC is also observable. 
The asymmetry in the onset of the gamma oscillations is clearly visible in the
continuous wavelet transform obtained from the experimental data and reported in Fig. 4G in \citep{butler2016}.
This asymmetry can be explained in an adiabatic framework by considering
the corresponding bifurcation diagram shown in Fig. \ref{FigPINGsub} (a). Indeed for the
sub-critical Hopf, the COs will emerge for $I_\theta > [H_c^{(e)} - H^{(e)}]$, but they will disappear for a different
value of the forcing, namely $I_\theta < [H_{SN}^{(e)} - H^{(e)}]$. Instead for a 
super-critical Hopf the emergence and disappearence of the oscillations will occur
at the same forcing amplitude, namely $I_\theta = [H_c^{(e)} - H^{(e)}]$.

\section*{Appendix B: Different PING set-ups}

\begin{figure*}[ht]
\begin{center}
\includegraphics[width=0.45\textwidth,clip=true]{fig14a.eps}
\includegraphics[width=0.45\textwidth,clip=true]{fig14b.eps}
\includegraphics[width=0.45\textwidth,clip=true]{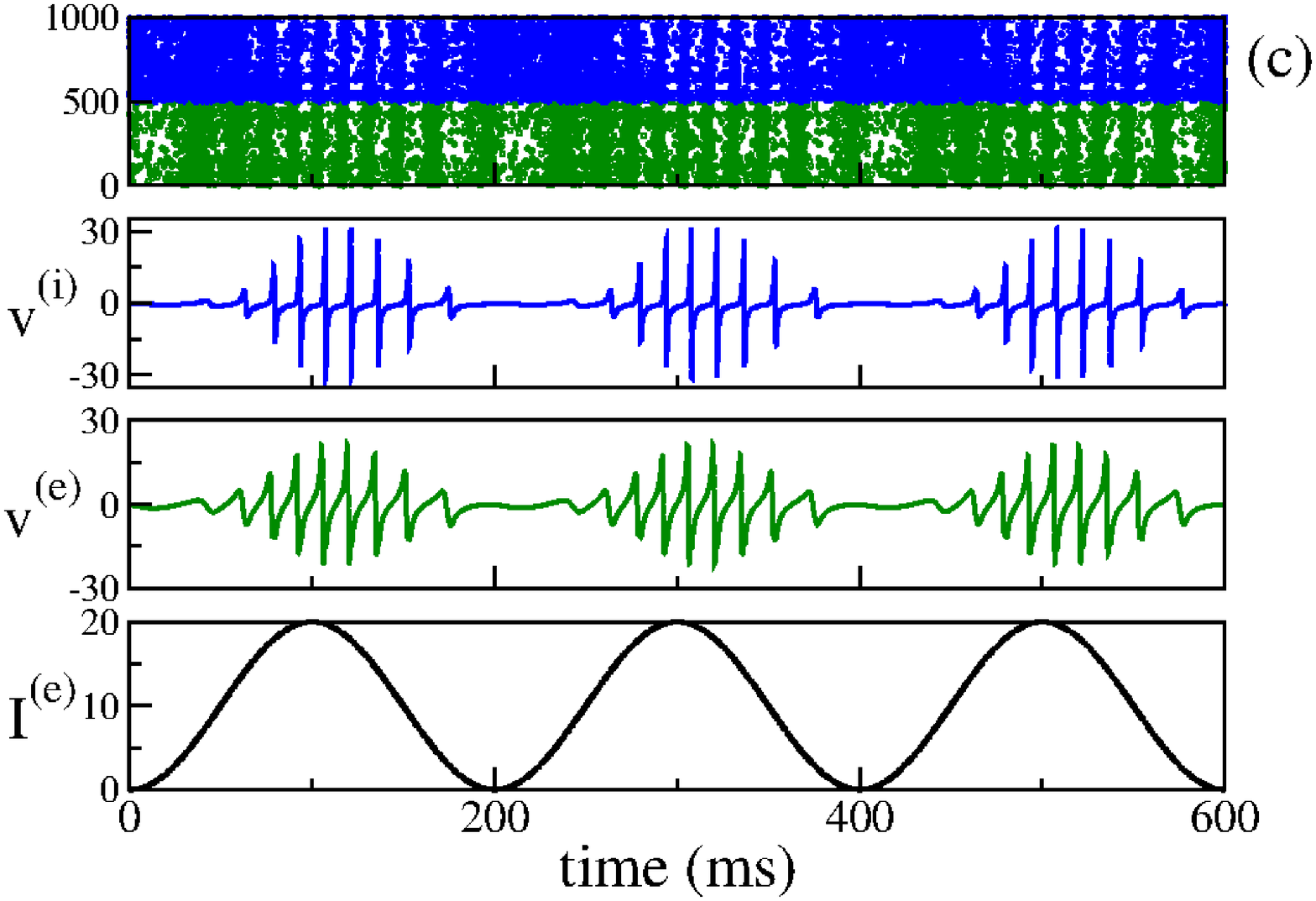}
\includegraphics[width=0.45\textwidth,clip=true]{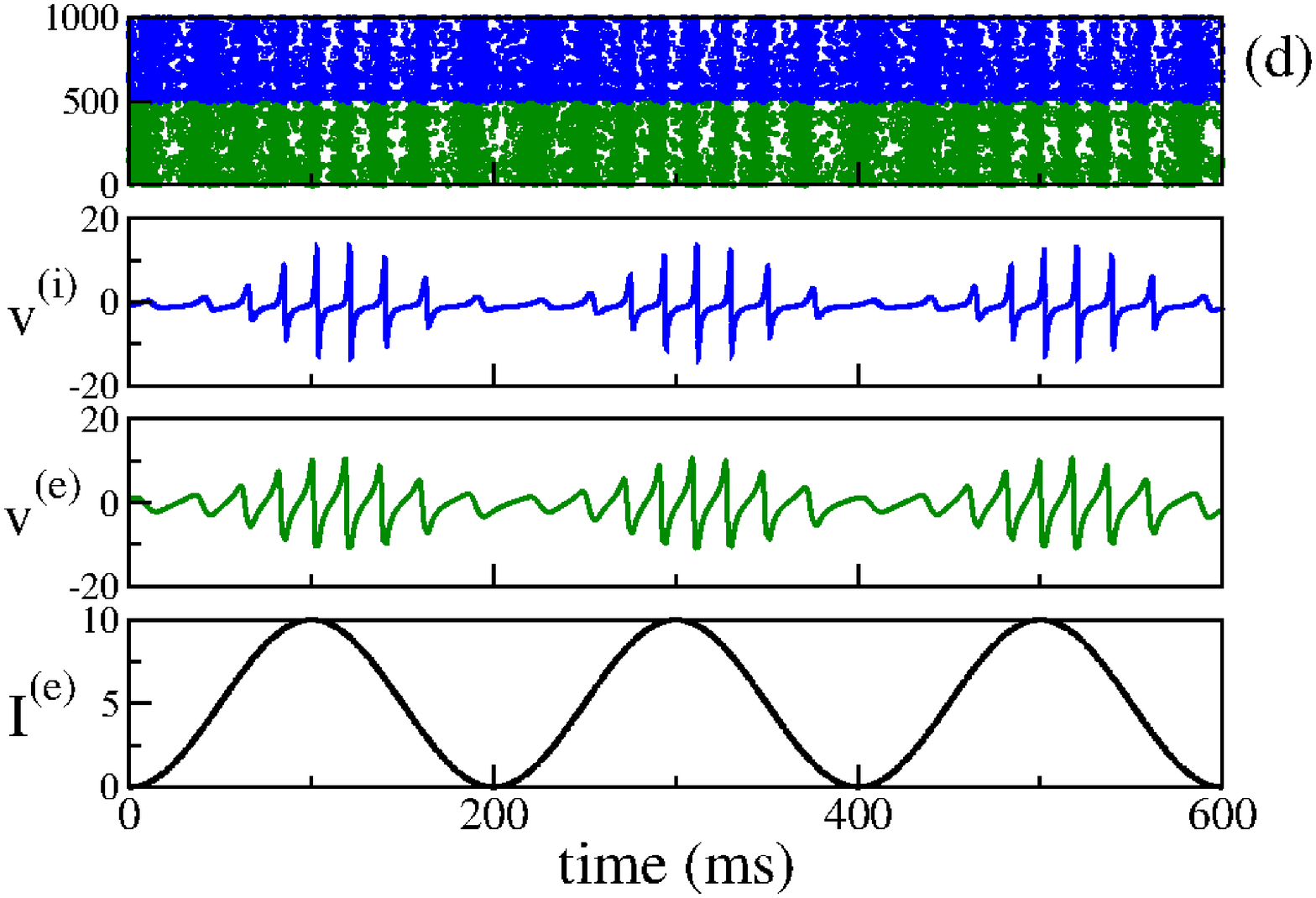}
\end{center}
\caption{ ({\bf Different PING set-ups}) Bifurcation diagram versus $H^{(e)}$ for the
PING$_{0}$ (a) and PING$_{I}$ (b) set-ups for $H^{(i)} = -0.5$.
In the corresponding insets, are reported the bifurcation diagrams as a function
of $H^{(i)}$, for $H^{(e)} = 10$. $\theta$-nested $\gamma$ oscillations emerging
in the PING$_0$ (c) and PING$_I$ (d) configurations for $I_0 = 20$ and $\nu_\theta=5$ Hz.
From top to bottom the raster plot where green (blue) dots refer to excitatory (inhibitory) neurons;
the average membrane potentials $v^{(i)}$ and $v^{(e)}$ as obtained by the evolution
of the neural mass models and the forcing currents $I^{(e)}$.
Parameters for the PING$_{0}$ set-up are $J_{ee} = J_{ii} = 0$,
while for PING$_{I}$ are $J_{ee} =0$ and  $J_{ii} = 8$. In both cases $J_{ie}=J_{ei}=10$
and $H^{(i)} = -0.5$. In the corresponding insets we set $H^{(e)}=10$.
}
\label{FigA1}
\end{figure*}

In the article we have considered an unique configuration giving rise to COs via the PING mechanism: namely, two cross
coupled inhibitory and excitatory populations with recurrent excitation and no recurrent inhibition
(i.e. $J^{(ee)} \ne 0$ and $J^{(ii)} = 0$). 
However, other network configurations can give rise to PING induced oscillatory regimes.
In particular, we have observed such oscillations with only cross-couplings in absence of recurrent excitation and inhibition
(i.e. $J^{(ee)} = J^{(ii)} = 0$), as well as in presence of recurrent inhibition only
(i.e. $J^{(ee)} = 0$ and $J^{(ii)} \ne 0$).
In the following we refer to the former configuration as PING$_{0}$ set-up, while the latter configuration with recurrent inhibition
is identified as PING$_{I}$ set-up. In both configurations the neural mass reproduces the emergence of $\gamma$ oscillations via a super-critical
Hopf bifurcation for increasing values of $H^{(e)}$, as shown in Figs. \ref{FigA1} (a) and (b).
Indeed the frequencies of the COs are in the range $[26:63.5]$ Hz ($[29.1:53.9]$ Hz) for PING$_{0}$
(PING$_{I}$) set-up.  In both configurations the corresponding bifurcation, as a function of the parameter $H^{(i)}$, is sub-critical and COs disappear for sufficiently
positive values of $H^{(i)}$, analogously to what reported in the main text for the PING set-up with only recurrent excitation. 
It should be stressed that the standard Wilson-Cowan neural mass model gives rise to COs
only in presence of a recurrent excitation \citep{wilson1972}, thus being unable to reproduce the spiking network dynamics \citep{dumont2019}.

In presence of an external $\theta$-forcing with $\nu_\theta=5$ Hz,
we clearly observe $\theta$-nested $\gamma$ oscillations, as shown in the raster plots 
reported in Figs. \ref{FigA1} (top rows of panels (b) and (d)). These oscillations are phase
amplitude modulated from the forcing, as it results to be evident from the shape of 
the mean membrane potentials $V^{(e)}$ and $V^{(i)}$ reported in 
the middle rows of Figs. \ref{FigA1} (c) and (d).

\bibliographystyle{frontiersinHLTH&FPHY}
\bibliography{biblio}

\end{document}